\documentclass[journal,twocolumn,10pt]{IEEEtran}

\usepackage[usenames]{color}

\normalsize

\usepackage{cite, amsmath,amsfonts,amsthm,amssymb}
\usepackage{algorithmic,algorithm,float,bbm}
\usepackage{multirow,balance,hyperref}
\usepackage{enumerate,bbding}
\usepackage{epsfig,graphicx,subfigure}

\theoremstyle{plain}

\theoremstyle{definition}

\theoremstyle{remark}

\usepackage{booktabs}
\usepackage{threeparttable}

\graphicspath{{images/}}

\usepackage{graphics}
\usepackage{array}
\usepackage{multirow}
\usepackage{adjustbox}

\usepackage{url}

\begin{document}


\title{Edge-Based Video Analytics: A Survey}

\author{\IEEEauthorblockN{Miao Hu\IEEEauthorrefmark{1}, Zhenxiao Luo\IEEEauthorrefmark{2}, Amirmohammad Pasdar\IEEEauthorrefmark{3}, Young Choon Lee\IEEEauthorrefmark{4}, Yipeng Zhou\IEEEauthorrefmark{5} and
Di Wu\IEEEauthorrefmark{6}}\\
\IEEEauthorblockA{School of Computer Science and Engineering and the Guangdong Key Laboratory of Big Data Analysis and Processing,
Sun Yat-sen University, China\IEEEauthorrefmark{1}\IEEEauthorrefmark{2}\IEEEauthorrefmark{6}\\
}
\IEEEauthorblockA{School of Computing, Macquarie University, Sydney, Australia\IEEEauthorrefmark{3}\IEEEauthorrefmark{4}\IEEEauthorrefmark{5}\\
\IEEEauthorrefmark{1}humiao@outlook.com,
\IEEEauthorrefmark{2}luozhx6@mail2.sysu.edu.cn, 
\IEEEauthorrefmark{3}amirmohammad.pasdar@hdr.mq.edu.au, \IEEEauthorrefmark{4}young.lee@mq.edu.au, \IEEEauthorrefmark{5}yipeng.zhou@mq.edu.au, 
\IEEEauthorrefmark{6}wudi27@mail.sysu.edu.cn}
}

\maketitle

\begin{abstract}
Edge computing has been getting a momentum with ever-increasing data at the edge of the network. In particular, huge amounts of video data and their real-time processing requirements have been increasingly hindering the traditional cloud computing approach due to high bandwidth consumption and high latency. 
Edge computing in essence aims to overcome this hindrance by processing most video data making use of edge servers, such as small-scale on-premises server clusters, server-grade computing resources at mobile base stations and even mobile devices like smartphones and tablets; hence, the term edge-based video analytics. However, the actual realization of such analytics requires more than the simple, collective use of edge servers. In this paper, we survey state-of-the-art works on edge-based video analytics with respect to applications, architectures, techniques, resource management, security and privacy. We provide a comprehensive and detailed review on what works, what doesn't work and why. These findings give insights and suggestions for next generation edge-based video analytics. We also identify open issues and research directions.

\end{abstract}

\section{Introduction}
\begin{figure*}
  \centering
  \includegraphics[width=0.65\textwidth]{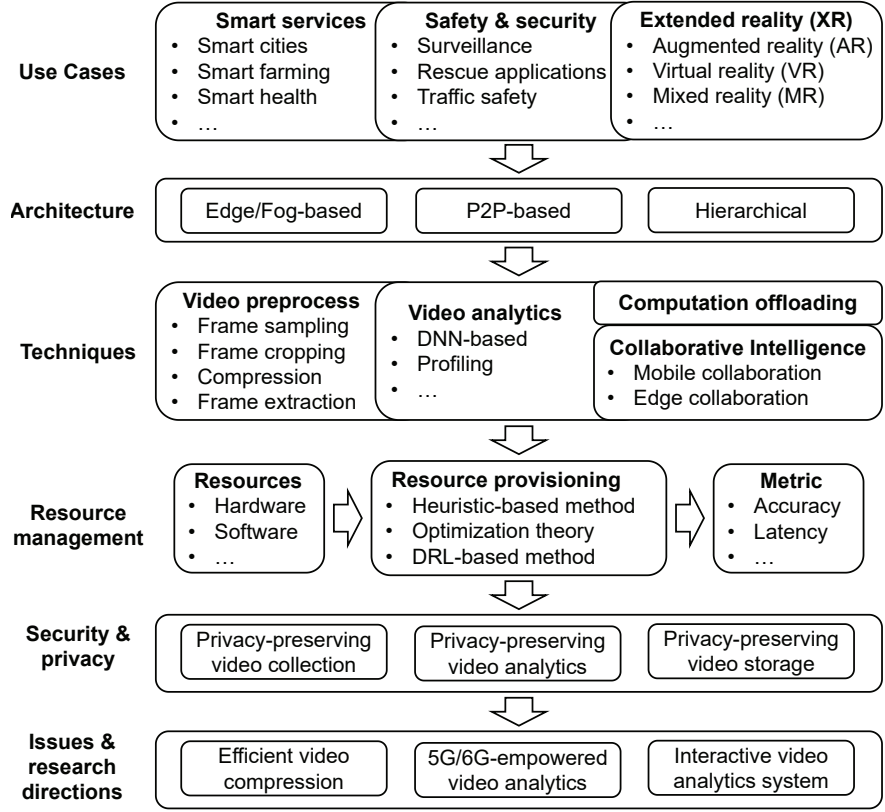}
  \caption{The structure of this article.}\label{structure}
\end{figure*}
In the past few decades, we have witnessed the explosive growth of data particularly with video data. Video cameras, such as closed-circuit television (CCTV) cameras, webcams and dashboard cameras, are everywhere. They are used for various purposes, such as surveillance, security and safety. 
They record ``everyone" and ``everything" all the time. In a 2015 Information Handling Services report~\cite{ihs2015}, authors estimated 245 million security cameras installed globally meaning one camera for every 29 people. However, it is not simply the video data these cameras produce, but insights from such data. In other words, the timely processing and analysis of such data is of great practical importance.

As reported by Fortune Business in Sights, the global video analyitcs market size is projected to reach USD\$ 12 billion by the end of 2026, exhibiting a compound annual growth rate of 22.67\% \cite{fortune}. Current video analytics applications are generally built on deep neural network models, which will bring great computational pressure from both model training and inference. Up until recently, these applications have run mostly in clouds~\cite{Yi2017SEC_LAVEA, Hung2018SEC}. However, the ``distant" clouds are facing serious challenges in meeting real-time processing requirements due to network bottleneck, i.e., high bandwidth consumption and high latency as video data has to travel back and forth over the internet.

Recently, the edge computing paradigm has emerged as a solution to that. It is rather a complementary approach to cloud computing making use of computing resources at the edge of the network, close to data sources. These resources are called edge servers. They include small-scale on-premises server clusters, server-grade computing resources at mobile base stations and even mobile devices like smartphones and tablets. While running video anayltics application at the edge with these computing resources cannot be perceived as a similar case to that in the cloud because there are some unique challenges in edge-based video analytics.

In this paper, we first provide a survey of edge-based video analytics applications and use cases. We then identify and discuss those challenges with a comprehensive survey of state-of-the-art works on edge-based video analytics. The following are four categories of these challenges.
\begin{itemize}
    \item \emph{Architecture Design}. Different architectures for edge-based video analytics may emphasize on various aspects and can be applied to different use cases and service scenarios.
    \item \emph{Video processing and analysis techniques}. Real-time processing requirements are a key challenge with edge servers that are often static with lower resource capacity compared to cloud servers. In-situ data analytics and collaborative processing are of particular interests for edge-based video analytics.
    \item \emph{Resource management}. With constraints on the quality of experience (QoE) metrics (e.g., accuracy, latency and energy consumption) and resources (hardware and software), it is of great importance to develop efficient resource management policies. It is largely unknown how effective traditional heuristic and optimization methods are for edge-based video analytics.
    \item \emph{Security and privacy}. The use of highly heterogeneous and decentralized edge servers is subject to security vulnerabilities. Besides, the processing, storage and caching of video data on these servers have serious privacy implications. 
\end{itemize}

While there have been several surveys on video analytics at the edge, they are not as comprehensive or in-depth as our survey in this paper. 
Shi \emph{et al.} \cite{Shi2016IEEEInternet} surveyed the works related to edge computing and its use cases. 
Zhou \emph{et al.} \cite{Zhou2019IEEE} summarized the research efforts on artificial intelligence (AI) from the perspective of edge computing and its corresponding intelligent applications. 
These works mainly discussed the technologies and applications with edge computing, however, the discussion on video analytics-related applications is limited.
Vega \emph{et al.} \cite{Vega2018IEEE} reviewed state-of-the-art QoE management methods for video-related services based on machine learning. Their goal is to optimize the QoE from the perspective of clients (i.e., devices), while efficiently utilizing network resources from the perspective of providers.
Barakabitze \emph{et al.} \cite{Barakabitze2019IEEE} studied QoE management solutions for edge-based multimedia applications. They mainly discussed about how to provide users with better video services. They considered this aspect as well as efficient video processing and analytics. 
As the most similar survey to our work, Jedari \emph{et al.} \cite{Jedari2020surveys} studied the state-of-the-art researches on edge video caching, edge computing, and communication. However, they mainly considered the combined use of resources at the edge to support several video-oriented applications, while only a small part of the content discussed the studies on video analytics. 

This paper provides a comprehensive and detailed review on what works, what doesn't work and why. These findings give insights and suggestions for next generation edge-based video analytics. We also identify open issues and research directions.

The structure of this paper is shown in Fig. \ref{structure}. In particular, Section~\ref{sec_scena} presents the general use cases and service scenarios for edge-based video analytics.
Section~\ref{sec_archi} summarizes the proposed architecture for edge-based video analytics.
Section~\ref{sec_tech} illustrates the technologies and methods for edge-based video analytics.
Section~\ref{sec_resource} reveals relation between resources and performance for edge-based video analytics, and discusses the scheduling strategies in resource-limited video analytics scenarios.
Section~\ref{sec_privacy} discusses the security and privacy issues occurred in edge-based video analytics system design.
Section~\ref{sec_issue} summarizes the challenging issues and outlooks future research directions.

\section{Use cases and service scenarios}\label{sec_scena}
This section introduces use cases and service scenarios built with edge-based video analytics.
They are divided into three categories: \emph{Smart Cities}, \emph{Safety and Security}, and \emph{XR (including AR, VR, and MR)}.

\subsection{Smart Services}

\subsubsection{Smart city}
The ``smart city" takes advantages of artificial intelligence into the urban construction for vehicle and human monitoring, city management and regulation of city flows and processes \cite{Jun2017SJ, Hu2018IOTJ}. Smart city applications include but are not limited to road traffic monitoring, road safety and security control, smart parking control, stolen cars search, etc. 
Specifically, Zhang \emph{et al.} \cite{Zhang2017nsdi} highlighted the resource-quality mapping correlation in smart city applications, including license plate reader, vehicle counter, crowd classifier and object tracker.
The core technology for most smart city applications is video analytics based on automated object detection \cite{Ananthanarayanan2019MobiSys}. However, it is tremendously high for the computation requirement for real-time object detection on live video streaming and the communication requirement for video uploading from cameras to the remote cloud server. Fortunately, it has been proved that the edge-based solution can help improve system efficiency and user experience when enjoying the smart city services. 
For example, Grassi \emph{et al.} \cite{Grassi2017SEC} proposed an edge-based in-vehicle video analytics system for monitoring parking spaces.
Xie \emph{et al.} \cite{Xie2018CyberC} proposed a video analytics-based intelligent indoor positioning system with the aid of edge computing.
With deep neural networks (DNNs), Barthelemy \emph{et al.} \cite{Barthelemy2019Sensors} designed an edge-based computer vision framework to monitor transportation while ensuring privacy of citizens. 
These works offer a novel way to analyze videos at edge to achieve bandwidth saving as well as data privacy.

\subsubsection{Smart farming}
The ``smart farming" utilizes the technology of artificial intelligence to improve the quantity and quality of crops. By precisely improving the development of intellectual and automatic agriculture machine, farmers can significantly increase the efficiency of agricultural cultivation. Video analytics techniques can offer farmers an effective way to monitor the status and requirements of their animals or crops and adjust the planting methods correspondingly, thereby preventing animal and crop diseases and enhancing their health \cite{smartfarm}. The agricultural automation also presents a high requirement on the application latency, where the edge-based farming-related video analytics applications are expected to be tremendously focused in the near future. Recently, Alharbi \cite{Alharbi2021Access} proposed an integrated edge-cloud architectural paradigm to enhance the energy-efficiency of smart agriculture systems and diminish carbon emissions.

\subsubsection{Smart health (e-Health)}
The public and personal health has always been the first place. In the year 2020, the World Health Organization declared the outbreak of COVID-19 as a pandemic.
To prevent the virus spread, it is a critically important issue to conduct a real-time temperature scanning on people in public areas and workplaces. 
Recently, many countries have deployed the AI thermal cameras for automated and contactless monitoring, especially for group temperature scanning. 
This can help implement strong protective measures while keeping the economy going. 

However, there are still some challenges for conducting efficient thermal video analytics. 
First and most important, it is essential for public health monitoring to run the accurate video analytics for temperature scanning and people tracking. 
Second, the temperature scanning results should be retrieved back in a real-time manner. 
Third, the cameras not only store temperature data but also personal identifiable thermal information, which might cause  privacy issues.
From the authors' perspective, the edge-based video analytics framework can better solve the above-mentioned issues than either the cloud-based video analytics and the in-situ video analytics. Specifically, a low latency performance can be achieved by placing the computation-intensive tasks (e.g., temperature scanning and people tracking) to the nearby edge servers. Meanwhile, the style of data storage at the edge brings information closer to the location, and it is tipped to improve privacy and security protection for  users.

\subsubsection{Smart business}
A key example for smart business is Amazon Go, which is a new kind of humanless stores without manual checkout.

Relied on computer vision techniques, the automated checkout system can map sales actions (e.g., picking products and checking out) to consumers, enabling the merchant to accurately charge customers for their picked products, e.g., \cite{Das2017CVPR}. In smart business system design, the video analytics technology has also been verified to be helpful in studies such as Cheng \emph{et al.} \cite{Cheng2018CoRR} and Xu \emph{et al.} \cite{Xu2019ISBN} where the former harnessed constrained resources in service industry via video analytics while the latter used autonomous cameras for object counting. 

Another popular business case is unmanned aerial vehicles (UAVs, also known as drones). Due to mobility and low cost, UAVs can help in various business applications such as express industry \cite{Wang2019IOTJ, Funabashi2019RTSS, Huang2020TITS}. Most drone-based delivery designs focused on the path scheduling problem aiming to minimize the total delivery time.

\subsubsection{Smart education}
Smart education enables learners to learn by accessing digital resources through Internet.
Long \emph{et al.} \cite{Long2018FIE} presented a video analytics-based lecture framework that transforms literal teaching contents into visual formats, based on the lecture video system deployed at the University of Houston. 
Jang \emph{et al.} \cite{Jang2018SEC} implemented a smart conference system with two prototype video analytics applications (monitoring and tracking).
Tarasov \emph{et al.} \cite{Tarasov2018AIST} addressed the emotion classification application by utilizing video analytics in education systems.
Hu \emph{et al.} \cite{Hu2019TPDS} first designed an edge-based framework for the video analytics assisted education system design in the real-time manner.
The edge-based video analytics-assisted smart education applications have also been verified in practical system designs, e.g., the high-performance computing (HPC) education platform \cite{Wu2020PAAP_VBSSR, Wu2020PAAP}.

\subsection{Safety and Security}

\subsubsection{Surveillance}
In major public safety events (e.g., terrorist attacks in public transportation scenarios), law enforcement may want to track down the identified perpetrators \cite{Zhang2015MobiCom}. Across modern cities, a large amount of cameras are installed in public areas around us, including underground transportation networks, ground transportation networks, and air transportation networks. It is urgently required that law enforcement and counter-terrorism departments can realize the real-time tracking on the public threats \cite{Schindler2019MMM}.

By placing cameras in public places such as roads, public transport, retail stores, parks and libraries, relevant departments can help prevent, track and solve crime problems via video analytics technologies \cite{Khochare2019CCGRID}.
Many works focused on the video analytics-based surveillance applications.
For example, Yi \emph{et al.} \cite{Yi2017SEC_LAVEA} highlighted the low-latency video analytics requirements for public safety applications, such as counter-terrorism.
Jain \emph{et al.} \cite{Jain2019HotMobile} focused on applications based on multiple cameras, like crowd control and spotlight search. 
To sum up, the ``real-time" characteristic is one of the most essentially focused metrics in surveillance applications, where the edge computing technology can help boost the performance.

\subsubsection{Rescue applications}
Different from surveillance applications, rescue applications need to not only detect, but also identify and track the corresponding objects. 
In response to a disaster, the capabilities of a remotely controlled drone to search large areas quickly and efficiently with high definition cameras make rescue operations much more efficient and effective.
Chowdhery \emph{et al.} \cite{Chowdhery2018SECON} proposed a novel approach for drone video analytics based on model predictive compression methods.
Wang \emph{et al.} \cite{Wang2018SEC} built an adaptive video analytics pipeline for searching tasks in domains such as life search and rescue. 
George \emph{et al.} \cite{George2019HotMobile} investigated the use of drones for live rescue, where the key technical challenge lies in the ingest of live video streams, based on which the architectural plan can be timely updated.

\subsubsection{Road and traffic safety}
Given cameras installed along highways and city streets, the video analytics technologies can be used to re-identify and track the suspect's vehicle \cite{Lu2016SoCC_Optasia}. 
Qiu \emph{et al.} \cite{Qiu2018IoTDI} designed and implemented a video analytics system, named Kestrel, that tracks the vehicles' trajectories with the aid of a heterogeneous camera network. 

For the case of traffic monitoring, Chen \emph{et al.} \cite{Chen2016SEC} proposed an edge-based video analytics system to timely get the vehicle speed information and track speeding vehicles.
For urban traffic surveillance, Chen \emph{et al.} \cite{Chen2016BigMM} proposed a dynamic video stream processing scheme with the ability of real-time video processing and decision making. 
For the purpose of monitoring the mobility within a network, Barthelemy \emph{et al.} \cite{Barthelemy2019Sensors} designed a sensor that can detect and track objects of interest in real-time video feed with the aid of video analytics technologies.

\subsection{XR (AR, VR, and MR)}
Extended reality (XR) technologies such as virtual reality (VR), augmented reality (AR), and mixed reality (MR) have emerged as methods to create simulated experiences that resemble the real world.
In the XR processing pipeline, a crucial task is the detection and tracking of real-world object positions, which enable accurate overlaying of virtual annotations on top of them \cite{Hu2021JNCA}.
For example, Jain \emph{et al.} \cite{Jain2015MobiSys_OverLay, Jain2016CoNEXT} proposed the first effort on end-to-end AR system implementation.
While commercial mixed reality platforms are capable of detecting surfaces and certain objects (e.g., a particular person) with the understand of 3D geometry of the scene, they often lack the capability to track and detect intricate and non-stationary objects.
In augmented vehicular reality system proposed in \cite{Qiu2018MobiSys_AVR}, vehicles exchange raw dynamic 3D sensor outputs (also named as point clouds).
Liu \emph{et al.} \cite{Liu2019MobiCom} designed a high accuracy object detection system for commodity AR/MR systems with 60fps. The system separates the rendering and offloading pipelines and employs a fast object tracking approach to ensure detection accuracy.

Zhang \emph{et al.} \cite{Zhang2017VR_ARNetwork} found that the latency lower bound to enable cloud-based mobile AR with acceptable QoE is around 250ms, which means that there exists room for further improvements.
Ran \emph{et al.} \cite{Ran2019HotNets} firstly provided a clear illustration of the information flow in multi-user augmented reality (AR). Specifically, they examined both Google ARCore\footnote{ARCore, a Google augmented reality SDK designed for Android system.} and Apple ARKit\footnote{ARKit, an Apple augmented reality SDK designed for iOS.}, and found that both of they employ either cloud-based or peer-to-peer architectures, where the edge-based architecture has not yet been taken into consideration. 
The battery capacity is the major constraint for executing XR applications.
Apicharttrisorn \emph{et al.} \cite{Apicharttrisorn2019SenSys} found that locally executing DNN-based object detection on mobile devices could significantly increase battery usage, which is a major concern for mobile users. They found that the screen, camera, and operating system already consume a considerable amount of battery (3-4W in their measurements), and the DNN executions can further drain a significant portion (1.7-3W) of the battery.

To further compensate for the lack of bandwidth and computing capability, Qiao \emph{et al.} \cite{Qiao2018InternetComputing_WebAR} proposed a web AR service-provisioning framework with edge servers. Moreover, the collaboration among edge servers can enhance the XR system performance. 
Zhang \emph{et al.} \cite{Zhang2018HotMobile} enabled the coordination and collaboration of computation resources, e.g., sharing the results of computation intensive AR tasks, and annotating high quality AR modifications by users.

\section{Architecture}\label{sec_archi}
This section specifies and compares several types of existing edge-based video analytics system architectures. Different architectures to the edge-based video analytics may prioritize different aspects and have distinct optimization approaches. In this regard, we describe the components and features of edge-based video analytics systems and provide examples of typical systems for each architecture type.

\subsection{Edge/Fog-based Architecture}
Edge computing or fog computing, as an extension of cloud computing, allows for computing tasks to be performed at the edge of the network with low latency and real-time computing capabilities. With the help of edge computing, video analytics tasks can be conducted on the edge servers instead of only being executed on end devices with relatively limited computing resources. The edge/fog-based architecture generally consists of two tiers, the end device tier and the edge computing tier, in which the near-site edge tier is essential for achieving  real-time video analytics. Generally, a wide range of devices can serve as edge nodes, e.g., smart phones, laptops and drones. The edge computing platforms (e.g., ParaDrop \cite{Liu2016SEC_ParaDrop}) provide application program interfaces (APIs) to manage  edge nodes as well as running edge services.

\begin{figure}[h]
\centering
  \includegraphics[width=0.45\textwidth]{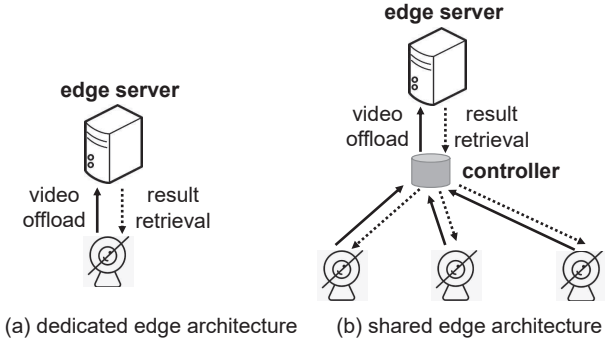}\\
 \caption{Two types of the edge/fog-based video analytics architecture.}\label{edge}
\end{figure}
Based on the operation modes, the edge-based architecture can be divided into two types, i.e., the dedicated edge-based architecture and the shared edge-based architecture.
As shown in Fig. \ref{edge} (a), each camera is equipped with a dedicated server for the dedicated edge-based architecture. While for the shared edge-based architecture, cameras will share the resource of an individual edge server as illustrated in Fig. \ref{edge} (b).

\subsubsection{Dedicated Edge-based Architecture}
The dedicated edge-based architecture is a simple and reliable way to build a video analytics system.
One example of an edge-based system that leverages real-time video surveillance is Vigil \cite{Zhang2015MobiCom}. It utilizes edge computing to allow for wireless video surveillance to scale to multiple cameras. The Vigil architecture is designed to enable basic vision analytics tasks to be performed at the edge nodes, which are connected to camera devices. To reduce the transmission overhead, only relevant portions of the video feed are uploaded to a controller. 
Grassi \emph{et al.} \cite{Grassi2017SEC} presented the ParkMaster architecture for road sign detection and utilized smartphone cameras with camera calibration for video processing. 
King \emph{et al.} \cite{King2020EdgeSum} designed the EdgeSum framework for video summarization and compression of dash cam (a.k.a. drive recorder) videos using mobile devices as edge servers before uploading to the cloud.

However, the number of edge servers will increase with the number of deployed cameras. With the increase of installed cameras, it will impose a rather heavy burden on installing too many edge servers. In view of  this challenge,  the shared edge-based architecture is preferred in many studies.

\subsubsection{Shared Edge-based Architecture}
Different from the dedicated architecture, the shared edge-based video analytics architecture is built on the virtualization technology.
For example, Jang \emph{et al.} \cite{Jang2018SEC} proposed an edge camera virtualization architecture that leverages an ontology-based application description model to virtualize the camera. They used container technology to decouple the physical camera and support multiple applications on board, thus improving resource utilization and flexibility in edge computing environments.
Wang \emph{et al.} \cite{Wang2017SmartIOT} proposed a smart surveillance system that leverages edge computing and application program interface technologies to enable flexible monitoring of security events in urban regions where have a dense network of cameras, with low latency and minimal backbone bandwidth consumption.
Similarly, a real-time video analytics system called EdgeEye~\cite{Liu2018EdgeSys_EdgeEye} was proposed. EdgeEye offered a high-level abstraction of partial video analytics functions through DNNs and provided tools to deploy and execute DNN models on edge servers.

Specially, most surveillance and rescue applications with cameras on  drones are realized on the shared edge server.
In \cite{Wang2018SEC}, the edge server is connected directly to the LTE base station and packets transmitted from the drones are directed to the edge server without traveling through the Internet backbone.
George \emph{et al.} \cite{George2019HotMobile} proposed a system architecture that leverages edge computing resources for drone-sourced video analytics in live building inspection. The high computational demands are met by using substantial edge computing resources, while the ability to virtualize on an edge server allows for the deployment of a virtual machine that contains the engineering and architectural drawings for the construction site.

\subsection{P2P-based Architecture}

In order to improve the analytics performance by utilizing cross-camera correlations, the analytics pipeline must have the capability to access inference results from related video streams and enable peer-triggered inference at runtime. It means that any relevant camera can assign an analytics task to process a video stream regardless of the time, which divide the logical analytics pipeline from its execution.

In order to achieve this, the inference results must be shared between pipelines in real-time. Although prior research has explored task offloading across cameras and between the edge and cloud, Jain \emph{et al.} \cite{Jain2019HotMobile} argued that the video streams of other related cameras should be considered in such dynamic triggering.

At present, the execution of video analytics pipelines is typically predetermined in terms of resource allocation and video selection. However, to leverage cross-camera correlations, a pipeline should have knowledge of the inference results of other relevant video streams and support real-time triggering based on this information. This enables the compute resources of related cameras to handle the analytic tasks dynamically according to the video streams.

Stone \emph{et al.} \cite{Stone2019SECON} proposed Tetris system that focuses on scalable video analytics at the edge. 
The system identifies active regions across all video feeds and compresses them into a compressed volume which are then passed a Convolutional Neural Networks (CNNs) layer and carefully organized system pipelines to achieve a high parallelism.
Luo \emph{et al.} \cite{Luo2018SEC_EdgeBox} proposed an EdgeBox solution that a group of cameras are managed by an edge device and deployed on the same local area network. This approach is suitable for covering relatively small areas. However, when edge nodes are connected and collaborate to perform complex activity detection utilizing deep learning and computer vision, they can cover larger areas such as a building or a factory.

\begin{table*}
\renewcommand\arraystretch{1.35}
\caption{Summary of the existing architectures for edge-based video analytics}\label{QoE_table}
\centering
\linespread{1}\selectfont
\begin{tabular}{|p{2cm}<{\centering}|p{2cm}<{\centering}|p{3cm}<{\centering}|p{3cm}<{\centering}|p{2.5cm}<{\centering}|p{3cm}<{\centering}|}
\hline \bf{Type} & \bf{Literature} & \bf{End Device Layer} & \bf{Edge/Fog Layer} & \bf{Controller} & \bf{Cloud Layer} \\
\hline \multirow{7}{*}{\shortstack{Edge/Fog-based \\ Architecture}} 
& Vigil \cite{Zhang2015MobiCom} & video recording and offloading & simple vision analytics & Internet & $\times$ \\
\cline{2-6} & Jang \emph{et al.} \cite{Jang2018SEC} & IoT camera which has specific functionalities (e.g., video recording) & edge-based video analytics & accepts video requests from applications via the cloud (or the user) & $\times$ \\
\cline{2-6} & Wang \emph{et al.} \cite{Wang2017SmartIOT} & video recording and offloading & edge-based video analytics & $\times$ & $\times$ \\
\cline{2-6} & EdgeEye \cite{Liu2018EdgeSys_EdgeEye} & video recording and offloading & an easy and efficient way to execute DNN models & ParaDrop & $\times$ \\
\cline{2-6} & Wang \cite{Wang2018SEC} & drone-sourced cameras & edge servers connected to LTE base stations & $\times$ & without traversing the Internet backbone \\
\cline{2-6} & George \cite{George2019HotMobile} & drone-sourced cameras & edge servers are used to meet the high computation and virtualization demands & $\times$ & $\times$ \\
\cline{2-6} & Dao \emph{et al.} \cite{Dao2017ICDCS} & extract and upload features when environmental changes are detected & installed in each camera & detection metadata collection & $\times$ \\

\hline \multirow{2}{*}{\shortstack{P2P-based \\ Architecture}} 
& Jain \emph{et al.} \cite{Jain2019HotMobile} & video recording and offloading & a video analytics pipeline about resources to use and video to analyze & $\times$ & $\times$ \\
\cline{2-6} & Tetris \cite{Stone2019SECON} & video feeds & a solution for large-scale video analytics in edge & identifies active regions across all video feeds and compresses them into a compressed volume & \checkmark \\

\hline \multirow{8}{*}{\shortstack{Hierarchical \\ Architecture}} 
& LAVEA \cite{Yi2017SEC_LAVEA} & cameras & clients are one-hop away from edge server via wire or wireless links & $\times$ & run heavy tasks on resource rich cloud node to improve response time or energy cost \\
\cline{2-6} & Anveshak \cite{Khochare2019CCGRID} & cameras & edge-based video analytics & automates application deployment and orchestration across edge and cloud & \checkmark \\
\cline{2-6} & Ali \emph{et al.} \cite{Ali2018ICFEC} & cameras & deep learning at the edge & deep learning across resources &  set to achieve an improved performance for object inference \\
\cline{2-6} & Ananthanarayanan \emph{et al.} \cite{Ananthanarayanan2017computer} & cameras & decode video, detect objects, and perform video analytics tasks & $\times$ & \checkmark \\
\cline{2-6} & Chen \emph{et al.} \cite{Chen2016SEC} & surveillance application layer &  process end users data and return the results back on time & $\times$ & \checkmark \\
\cline{2-6} & Perala \emph{et al.} \cite{Perala2018ISCAS} & capture videos in different resolution based on their configuration & computing devices directly connected to the cameras & the gateway on which the devices are connected and share & the cloud server to which the gateway is connected \\
\cline{2-6} & Drolia \emph{et al.} \cite{Drolia2017ICDCS_Cachier} & cameras & use an edge server as a cache with compute resources & $\times$ & over the Internet\\
\cline{2-6} & CloudSeg \cite{Wang2019HotCloud} & camera sensor & send a downsampled high-resolution video adaptively over Internet & $\times$ & process the video with DNN inference, and return the inference results to the edge \\

\hline
\end{tabular}
\end{table*}

\subsection{Hierarchical Architecture}

\begin{figure}[h]
\centering
  \includegraphics[width=0.45\textwidth]{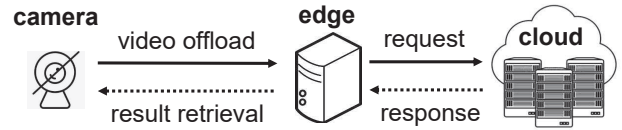}\\
 \caption{The three-layer hierarchical video analytics architecture consists of camera layer (also called user layer), edge layer, and cloud layer.}\label{edge-cloud}
\end{figure}
It is a challenging task to coordinate highly heterogeneous computing nodes to work as homogeneous computing nodes, which is shown in Fig. \ref{edge-cloud}. The three-layer hierarchical video analytics architecture consists of an end device layer (also called application layer or user layer), an edge layer, and a cloud layer. The camera, edge, and public cloud clusters differ in the available hardware types. For instance, GPUs are commonly found in some clusters (including the cameras), while other types of hardware, such as field-programmable gate arrays (FPGAs) and application-specific integrated circuits (ASICs), are typically found in public clouds \cite{Ananthanarayanan2017computer}.

FilterForward \cite{Canel2019SysML} is a platform that enables multi-tenant video filtering for edge nodes with limited bandwidth. While purely edge-based approaches are limited by static compute and storage resources, datacenter-only analytics require heavy video compression for transport. It addresses these challenges by allowing applications to split the work flexibly into edge and cloud. By leveraging high-fidelity data available at the edge, this approach allows for relevant video sequences to be made available in the cloud.

Similarly, Yi \emph{et al.} \cite{Yi2017SEC_LAVEA} utilized the cloud layer to execute computationally intensive tasks on powerful cloud nodes in order to reduce response time or improve energy efficiency.
Khochare \emph{et al.} \cite{Khochare2019CCGRID} built Anveshak framework, which automates the application deployment and orchestration across edge and cloud resources.
Ali \emph{et al.}  \cite{Ali2018ICFEC} proposed a deep learning pipeline that utilizes resources at the edge, in-transit, and cloud to achieve low latency, reduced bandwidth costs, and improved performance of video analytics tasks.
Ananthanarayanan \emph{et al.} \cite{Ananthanarayanan2017computer} proposed a hierarchical geo-distributed infrastructure which consists of edge clusters and private clusters with heterogeneous hardware for video decoding, object detection, and other video analytics tasks.
Ran \emph{et al.} \cite{Ran2018INFOCOM_DeepDecision} proposed a framework that integrates front-end devices with more powerful backend ``helpers" (such as home servers) to enable local or remote execution of deep learning in the edge/cloud. 
In \cite{Chen2016SEC}, the system comprises of three different layers, where the edge layer is made up of different on-site smart devices that act as both data producers and computing nodes.
Besides the above mentioned works, most edge-based video analytics systems, e.g., \cite{Perala2018ISCAS, Drolia2017ICDCS_Cachier, Wang2019HotCloud, Guo2019TMM, Zhang2016SEC, Long2017TMM}, consist of three parts, including end devices, edge servers, and the cloud. To  authors' best knowledge, the hierarchical architecture has been verified as the most efficient and scalable one for edge-based video analytics.

\section{Techniques and Methods}\label{sec_tech}


In this section, we overview various technologies and methods for edge-based video analytics, describe their roles and functions, and list some typical examples for each type.

\subsection{Video Preprocessing}\label{subsec_preprocess}
Generally, the practice of offloading all raw videos to the edge or cloud will cause a huge burden on the network and may lead to intolerable latency. Facing this challenge, a direct approach is to preprocess the raw videos before offloading them to the edge or cloud. Many researches have applied various techniques for video preprocessing, including \emph{frame sampling}, \emph{frame cropping}, \emph{compression}, and \emph{feature extraction}. We briefly introduce these techniques as follows.

\begin{table*}
\renewcommand\arraystretch{1.35}
	\caption{Summary of the Video Preprocessing methods}\label{video preprocessing}
	\centering
\linespread{1}\selectfont
		\begin{tabular}{|p{2.5cm}<{\centering}|p{2cm}<{\centering}|p{2cm}<{\centering}|p{3cm}<{\centering}|p{3cm}<{\centering}|p{3cm}<{\centering}|}
			\hline \bf{Category} & \bf{Literature} & \bf{Method} & \bf{End Device Layer} & \bf{Edge/Fog Layer} & \bf{Cloud Layer} \\ \hline
			\hline \multirow{5}{*}{\shortstack{Frame \\ Sampling}} & Xu \emph{et al.} \cite{Xu2019ISBN} & Frame filtering & Split-phase planning for making heterogeneous sampling decisions & $\times$ & $\times$ \\
			\cline{2-6} &  Zhang \emph{et al.} \cite{Zhang2018ICPP} & Frame filtering & Apply multi-stage filters and a global feedback-queue mechanism & $\times$ & $\times$  \\
			\cline{2-6} & Zhang \emph{et al.} \cite{Zhang2016SEC} & Frame filtering & $\times$ & Use OpenCV & $\times$ \\
			\cline{2-6} & Chowdhery \emph{et al.} \cite{Chowdhery2018SECON} & Frame filtering & Use predicted drone trajectory & $\times$ & $\times$ \\
			\cline{2-6} & Wang \emph{et al.} \cite{Wang2018SEC} &  Frame filtering &  Transfer learning on the drone & $\times$ & $\times$  \\
			\cline{2-6} & \cite{Zhang2015MobiCom,Dao2017MASS} &  Cross cameras &  Jointly consider views from multiple cameras & $\times$ & $\times$  \\

			\hline \multirow{2}{*}{\shortstack{Frame \\ Cropping}} & Chen \emph{et al.} \cite{Chen2016SEC, Chen2016BigMM} & RoI-based  & $\times$ & Extract the RoI of frames & $\times$ \\
			\cline{2-6} & Guo \emph{et al.} \cite{Guo2019TMM} &  RoI-based & Apply a RoI based image compression algorithm & $\times$ & $\times$ \\

			\hline \multirow{3}{*}{\shortstack{Compression}} & Wang \emph{et al.}\cite{Wang2016TMM} &  Coding optimization  & $\times$ & Use Gaussian low-pass filtering & $\times$ \\
			\cline{2-6} & CloudSeg \cite{Wang2019HotCloud} &  Super resolution & $\times$ & Lower the quality of videos &  Run the SR procedure to reconstruct the high-quality videos \\
			\cline{2-6} & SR360 \cite{Chen2020NOSSDAV} &  Super resolution & Run the SR procedure to boost the video tiles & $\times$ &  Stream requested video contents to the client \\
			
			\hline \multirow{3}{*}{\shortstack{Feature \\ Extraction}} & Canel \emph{et al.} \cite{Canel2019SysML} &   DNN-based  & $\times$ &  Feature maps are produced from the intermediate results of a single reference DNN & $\times$ \\
			\cline{2-6} & George \emph{et al.} \cite{George2019HotMobile} &  Computer vision-based & Register the drone view to a reference image & Use SIFT matching to find correspondences between the live camera view and the reference images &  $\times$ \\
			\cline{2-6} & Jiang \emph{et al.} \cite{Jiang2018ATC} &  DNN-based & $\times$ & Exploit partial-DNN compute sharing among applications trained through transfer learning & $\times$ \\
			\hline
	\end{tabular}
\end{table*}

\subsubsection{Frame Sampling} \label{subsubsec_fs}

Frame sampling skips frames that may be ``useless'' in video streams for analytics tasks \cite{Lu18INFOCOM, Lu18TMC_CrowdVision}.
For example, Xu \emph{et al.} \cite{Xu2019ISBN} proposed a split-phase planning mechanism for making frame sampling decisions and resolving the tension of frame capturing/processing.
The FFS-VA system proposed in \cite{Zhang2018ICPP} utilizes two prepositive stream-specialized filters and a tiny-YOLO model \cite{tyolo} to filter out irrelevant frames.
Zhang \emph{et al.} \cite{Zhang2016SEC} extracted valuable information by using OpenCV\cite{pulli2012real} on edge nodes instead of sending the raw videos to cloud.
Chowdhery \emph{et al.} \cite{Chowdhery2018SECON} proposed a method that uses the predicted trajectory of a drone to choose and send the most relevant frames to a ground station, with the goal of maximizing the utility of application while minimizing the consumption of bandwidth.
Wang \emph{et al.} \cite{Wang2018SEC} leveraged state-of-the-art deep neural networks (e.g., MobileNet \cite{howard2017mobilenets}) with transfer learning technology to transmit the interesting data selectively from the video stream.
These solutions significantly reduce the number of video frames sent to the cloud for video analytics tasks.

Recently, some researches used the correlations among cross-cameras views to further improve the efficiency of frame sampling. 
For instance, Vigil \cite{Zhang2015MobiCom} only uploads the most relevant frames to the cloud accoding to the user's query when there are different views of an object or person captured by multiple cameras.
Similarly, Dao \emph{et al.} \cite{Dao2017MASS} utilized the joint camera views of the object to determine the views which achieve the best accuracy regard to the object of interest and improve the quality of detection. 
Collaboration between cameras reduces the amount of offloaded data, thereby reducing the bandwidth consumption.

\subsubsection{Frame Cropping}\label{sec_frame_crop}
Different from frame sampling, frame cropping focuses on regions of interest (RoIs) in video frames, such as the face regions for face detection tasks or the vehicle regions for vehicle monitoring tasks. 
Frame cropping removes the unimportant regions of the raw images, thus, reduces the data transferred in the network.
Chen \emph{et al.} \cite{Chen2016SEC, Chen2016BigMM} extracted the RoI containing the suspicious vehicle instead of whole video frame to improve the video analytics efficiency. Guo \emph{et al.} \cite{Guo2019TMM} captured image data on end devices in a real-time manner and compressed them by a RoI based image compression algorithm. The compressed images were then transmitted to the cloud or the edge server, reducing the bandwidth usage and improving the transmission efficiency.

\subsubsection{Compression}
The compression technology has been widely used in video analytics applications to ensure a high processing efficiency.
Wang \emph{et al.}\cite{Wang2016TMM} proposed an adaptive approach for compressing screen content videos based on their utility, which involves identifying and processing low-utility content with a Gaussian low-pass filtering. 
Rippel \emph{et al.}\cite{Rippel2019ICCV} developed an architecture for video compression that extends motion estimation to perform learned compensation beyond simple translations.
Fouladi \emph{et al.}\cite{Fouladi2017nsdi} presented a video encoder that reach a fine-grained parallelism while keeping compression efficiency.

Recently, the super resolution (SR) technology has been regarded as an essential solution for video compression and transmission.
With SR techniques, the cameras or edge servers can send the video stream in low resolution, while the high-resolution frames can be recovered from the low-resolution stream. 
Through SR technology, it can reduce the transmission delay and reduce the pressure on network bandwidth while meeting the high requirements of video quality and analytics accuracy \cite{Wang2019HotCloud, Chen2020NOSSDAV}.

For instance, Wang \emph{et al.} \cite{Wang2019HotCloud} presented CloudSeg, which reduces the quality of the video during transmission to the cloud, but then performs the super-resolution process on the cloud server to reconstruct high-quality videos before executing video analytics.
Yang \emph{et al.} \cite{Yang2018Access} proposed an SR method to minimize SR errors by dividing the training samples into multiple clusters and learning dictionaries to achieve more faithful reconstructions in edge video analytics.
Chen \emph{et al.} \cite{Chen2020NOSSDAV} presented SR360 framework, in which the low resolution video tile in 360-degree videos can be upsampled at the client side to a high resolution tile with SR techniques.
Guo \emph{et al.} \cite{Guo2019sensors} proposed a semantic-aware SR transmission system for wireless multimedia sensor networks. The system encodes different bit-rate video with semantic information on the multimedia sensor, and uploads them to users. On the user side, the video quality is enhanced using SR techniques.

\subsubsection{Feature Extraction}
Feature extraction is a process that extracts image information and identifies whether each image point belongs to an image feature or not. Based on a preliminarily trained and lightweight neural network, the extracted feature map can be directly sent to the video analytics functions at the edge or cloud server.
The extracted feature map can be regarded as inputs to the object classifier, or stored and converted into useful information for other functions \cite{Uddin2019Symmetry}. 

Canel \emph{et al.} \cite{Canel2019SysML} presented FilterForward, where the feature maps are extracted from video frames on edge servers by a single reference DNN. Micro-classifiers are trained to take these feature maps as input and return the relevance of the specific applications and the frames. Besides, FilterForward allows micro-classifiers to get the feature maps of one layer of the model, making it versatile enough to support various tasks.

George \emph{et al.} \cite{George2019HotMobile} proposed an edge-based prototype that employed computer vision algorithms for live building inspection with drone-sourced video. Their system used visual features and Scale Invariant Feature Transform (SIFT) features \cite{lowe2004distinctive} matching to identify relevance between the reference images and view of live camera.

Grassi \emph{et al.} \cite{Grassi2017SEC} presented ParkMaster, which captures the parked vehicles in the mobile camera, extracts the feature, and then uploads the data to the cloud to run a clustering algorithm to count the number of parked cars on the road.

Mainstream \cite{Jiang2018ATC} is an edge system for video processing that employs transfer learning from a common base DNN model to train multiple applications.  By sharing partial-DNN compute among these applications, it reduce the computing time of per-frame aggregation.
Kang \emph{et al.} \cite{Kang2017PVLDB_NoScope} performed model specialization by using the full neural network to generate labeled  training  data  and  subsequently  training  smaller neural networks that are tailored to a given video stream and to a smaller class of objects.

\subsection{Video Analytics}\label{subsec_DNN}

\begin{table*}
\renewcommand\arraystretch{1.35}
	\caption{Summary of the Video Analytics methods}\label{video analytics}
	\centering
\linespread{1}\selectfont
		\begin{tabular}{|p{2cm}<{\centering}|p{2cm}<{\centering}|p{2.5cm}<{\centering}|p{2.5cm}<{\centering}|p{2cm}<{\centering}|p{2cm}<{\centering}|p{2cm}<{\centering}|}
			\hline \bf{Category} & \bf{Literature} & \bf{Contribution} & \bf{End Device Layer} & \bf{Edge/Fog Layer} & \bf{Cloud Layer} & \bf{Evaluation}\\ \hline
			\hline \multirow{4}{*}{\shortstack{Deep Neural \\ Network-based \\ Processing}} & EdgeEye \cite{Liu2018EdgeSys_EdgeEye} & Provides a high-level abstraction of video analytics functions based on DNNs & $\times$ & Offload the live video analytics tasks to the server using its API & $\times$ & Inference Speed: 55 Mean FPS \\
			\cline{2-7} &  Augur\cite{Lu17MM} & A CNN performance analyzer to determine the efficiency of a CNN on a given mobile platform & Take a CNN configuration as the input and predicts the computational time and resource utilization of the CNN & $\times$ & $\times$ & Deploy 4 kinds of CNNs on mobile platforms to measure and analyze the performance and resource usage \\
			\cline{2-7} & NetVision \cite{Lu2020ToN_NetVision} & Leverages DNNs across a network of mobiles and edge devices to minimize the querying response time & Determine a sequence with minimizing the query response time for the video offloading and transmission. & $\times$ & $\times$ & The greedy algorithm is close to the optimum, and the adaptive algorithm performs better \\
			\cline{2-7} & Huang \emph{et al.} \cite{Huang18MM_QARC} & Leverages a deep reinforcement learning algorithm to achieve higher perceptual quality rate & $\times$ & DRL model selects sending bitrate for future video frames & $\times$ & Average video quality: 18\% - 25\% , decreasing in average latency with 23\% -45\% \\
			
			\hline \multirow{2}{*}{\shortstack{Profiling}} & VideoStorm \cite{Zhang2017nsdi} &  A video analytics system that processes thousands of video analytics queries on live video streams over large clusters  & $\times$ & $\times$ & An offline profiler generating query resource-quality profile and an online scheduler for resource allocation & Quality of real-world queries: 80\% improvement, lag: 7$\times$ better \\
			\cline{2-7} & VideoEdge \cite{Hung2018SEC} &  Achieve a trade-off between resources and accuracy for live video analytics & Merge common components across queries & Narrows the search space by identifying a ``Pareto band" of promising configurations & $\times$ & Improves accuracy: 5.4$\times$ - 25.4$\times$ \\
			\hline
	\end{tabular}
\end{table*}

\subsubsection{DNN-Based Processing}
Undoubtedly, DNNs contribute significantly to video analytics as they are believed to be the backbone of recognition and detection in images and videos. These DNNs can be a part of tool or library or they can be embedded into the hardware for on-device video analytics. 

Augur \cite{Lu17MM} is a tool providing insights about the efficiency of a CNN on specify mobile platform. Augur profiles and models the resource requirements of CNNs by using a configuration of the CNN to predicts the computational overhead and resource utilization of the model. The CNNs are selected from wide range of libraries such as ResNet \cite{he2016identity}, VGG \cite{simonyan2014very}, NASNet \cite{zoph2018learning}, AlexNet \cite{alexnet}, and GoogleNet \cite{gglnet} on NVIDIA TK1 and TX1 hardware \cite{jetsondevkit}. 

Augur \cite{Lu17MM} evaluates each model on CPU and GPU considering memory where holds the parameters of the CNN, stores intermediate data, and the workspace for computation. Unlike workstations, in mobile platforms GPU shares the system memory with CPU that is not addressed by Caffe, hence, it causes generating redundant copy of memory.
Augur finds that the matrix multiplications of CNN computation is the core for performance evaluation. They are measured by the BLAS and cuBLAS libraries for matrix multiplications on CPUs and GPUs, respectively. 
The matrix size of a CONV or FC layer is related to the input dimension (e.g., images size) and network configuration (e.g., kernel size). Besides, time measurements have shown that \emph{matmul} contributes more than 60\% of the computation time of a CNN on mobile platforms. To model the time for prediction purposes, several matrix sizes are benchmarked with respect to the number of kernels, the size of a kernel, and the spatial size of output feature maps. 
Therefore, Augur first parses the CNN descriptor and determines the minimal memory needed based on the type and setting of each layer. Then, Augur extracts matrix multiplications (matmuls) from the computation of the CNN and calculates the compute time of individual matmuls. Finally, Augur sums up the compute time of all matmuls to provide an estimate of the total computation time of the model on the mobile platform.

Deep learning-based video analytics systems may consist a few hyper-parameters, including \emph{learning rate}, \emph{activation function} and \emph{weight parameter initialization}.
Yaseen and colleagues \cite{Yaseen2019TSMC} addressed the challenge of optimizing hyper-parameters in deep learning-based video analytics systems. They proposed a mathematical model to evaluate the impact of different hyper-parameter values on system performance. Their work also included proposing an automatic object classification pipeline for efficient large-scale object classification in video data.
Nvidia Deep Learning GPU Training System (DIGITS) proposed a general DNN architecture.
Liu \emph{et al.} \cite{Liu2018EdgeSys_EdgeEye} used Nvidia Deep Learning GPU Training System (DIGITS), a general DNN architecture, to design their DetectNet, and proposed a framework called EdgeEye for video analytics in real-time at the edge. The EdgeEye server allows applications to offload live video analytics tasks through its API, eliminating the need for deep learning framework specific APIs.

Other usage of DNNs are found in \cite{Lu2020ToN_NetVision, Huang18MM_QARC, Han2016MobiSys} studies for reducing the response time and improving the transmission rates. Lu \emph{et al.} \cite{Lu2020ToN_NetVision} present NetVision  as a system for on-demand video processing that uses deep learning to minimize query response time across a network of mobile and edge devices.

The transmission rate is also improved in \cite{Huang18MM_QARC} which relies on a deep reinforcement learning algorithm focusing on higher perceptual quality rate with lower bitrate. The algorithm works based on a trained neural network to predict future bitrates based on observation of network status. The proposed model in their study relies on two separate neural networks; (1) it  precisely predicts future video qualities based on the previous video frames, and (2) a reinforcement learning algorithm to determine the proper bit rates with respect to the output of the first neural network. The output of the first neural network model relies on a combination of CNNs to extract image features and a recurrent neural network for capturing temporary features for providing the better video qualities. 

Han \emph{et al.} \cite{Han2016MobiSys} studied the DNNs usage to execute multiple applications on cloud-connected mobiles to process a stream of data. It relies on a trade-off between resource usage and accuracy to be coped with workloads considering less accurate variants of optimized models. Hence, an adaptive framework was presented to select model variants at different accuracy levels while staying with the request resource constraints and energy constraints. The framework keeps track of accuracy, energy usage, and resource usage to form a catalogue based on a series of model optimization techniques. It uses different settings and heuristically allocates resources in proportion according to the frequency of use and selection of the most accurate corresponding model variant. This selection is interpreted as a model execution either on the device or on the cloud such that which application model should be chosen at a specific time step and which models should be evicted from the mobile cache.

\subsubsection{Profiling}
Profiling reduces computation overload of a system for obtaining handful configurations for video analytics. VideoStorm \cite{Zhang2017nsdi} and VideoEdge \cite{Hung2018SEC} are systems that take advantage of profiling for improving performance. In Zhang \emph{et al.} \cite{Zhang2017nsdi}, the system processes many video analytics queries on live video stream over large clusters. The system leverages an offline profiler generating a profile of query resources and uses an online scheduler for resource allocation to maximize the performance in terms of both quality and response time. The quality and lag are encoded as utility functions in which becomes penalized for violations. The profiler uses greedy search and domain sampling specific for obtaining a set of handful configurations on the Pareto boundary of profile to be taken into account by the scheduler. 

Hung \emph{et al.} \cite{Hung2018SEC} showed that VideoEdge is capable of achieving a trade-off between resource consumption and model accuracy by narrowing down the configuration space in the hierarchical structure (i.e., cameras, private clusters, and public clouds) for live video analytics. In VideoEdge, the configuration space is downsized through maximum computation of maximum demand to capacity ratio for each configuration to find the dominant demand. This enables system to compare configurations across demand and accuracy. The system also leverages a profiler for efficient merging components (e.g., decoder) for improving the accuracy and reducing the computation workload.

\subsection{Computation offloading}\label{subsec_offload}

\begin{table*}
\renewcommand\arraystretch{1.35}
	\caption{Summary of the Computation offloading and Collaborative Intelligence methods}\label{CoCI}
	\centering
\linespread{1}\selectfont
		\begin{tabular}{|p{2cm}<{\centering}|p{2cm}<{\centering}|p{2cm}<{\centering}|p{2.5cm}<{\centering}|p{2cm}<{\centering}|p{2cm}<{\centering}|p{2cm}<{\centering}|}
			\hline \bf{Category} & \bf{Literature} & \bf{Method} & \bf{End Device Layer} & \bf{Edge/Fog Layer} & \bf{Cloud Layer} & \bf{Performance}\\ \hline
			\hline \multirow{4}{*}{\shortstack{Computation\\offloading}} & DeepDecision \cite{Ran2018INFOCOM_DeepDecision} & Compute locally or offload & Determine an optimal offload strategy by using estimates of network conditions, the application’s requirements and specific tradeoffs of models & Receive the offloaded tasks and process with big CNN& $\times$ & Improve accuracy: about 30\%\\
			\cline{2-7} &  PicSys\cite{Felemban2020TMC} & Compute locally or offload & Multi-stage decision technique to decide process the image locally with the accelerated CNN or offload & $\times$ & Receive the offloaded tasks and process with a more complex and accurate CNN & Time savings: 35\% \\
			\cline{2-7} & Neurosurgeon \cite{Kang2017ASPLOS} & Partition the entire task to each computing nodes & Intelligently partition DNN computation between the mobile and cloud, the mobile process parts of the DNN & $\times$ & Receive the intermediate data and process the rest parts & End-to-end latency: 3.1$\times$ improvement, mobile energy consumption: 59.5\% reduction \\
			\cline{2-7} & Emmons \emph{et al.}\cite{Emmons2019HotEdgeVideo} & Partition the entire task to each computing nodes & Process partly subject to a limit on computation & $\times$ & Receive the intermediate values limited by network capacity, and process further DNN inference. & $\times$ \\
			
			\hline \multirow{4}{*}{\shortstack{Collaborative\\Intelligence}} & Chameleon \cite{Jiang2018SIGCOMM_Chameleon} &  Mobile Collaboration  & Use temporal and spatial correlation of cross-camera to picks the best configurations for existing NN-based video analytics pipelines & $\times$ & $\times$ & 20-50\% higher accuracy with the same amount of resources, or the same accuracy with 30-50\% of the resources \\
			\cline{2-7} & Dao \emph{et al.}\cite{Dao2017ICDCS} &  Mobile Collaboration & Coordinate among cameras to deliver highly accurate detection of objects efficiently & $\times$ & $\times$ & Reduces energy consumption: 40\% \\
			\cline{2-7} & Long \emph{et al.}\cite{Long2017TMM} &  Edge Collaboration & $\times$ & Optimally form devices into video processing groups and dispatch video chunks to proper groups & $\times$ & Improve the human detection accuracy: 19\% \\
			\cline{2-7} & Wang \emph{et al.}\cite{Wang2017SmartIOT} &  Edge Collaboration & $\times$ & A group of VMs/VNFs launched to work together for a surveillance task & $\times$ & Latency: about 6.0$\times$ improvement \\
			\hline
	\end{tabular}
\end{table*}

Deep learning algorithms are often computationally intensive, while the front-end equipment usually lacks the computing power for executing large-scale deep learning tasks. Transferring the data to a powerful cloud and executing deep learning algorithms in the cloud may cause unacceptable latency for users. Cloud-based solutions for deep learning are dependent on reliable network access, and it is challenging to offload the full or partial compute tasks to the proximate edge servers.

\subsubsection{Full Offloading}
Some researches focused on the offloading strategy to decide whether to perform lightweight analytics on the edge or send the videos or images to the cloud for more computationally intensive analytics.
By considering current network conditions and application’s requirements as the trade-offs, Ran \emph{et al.} \cite{Ran2018INFOCOM_DeepDecision} presented DeepDecision to determine an optimal offload strategy in real-time AR applications, and decide either analyze the input video locally with small CNNs or send to the server to analyze with a big CNN.
Felemban \emph{et al.} \cite{Felemban2020TMC} proposed PicSys, an intelligent system that decides whether to process images locally with an accelerated CNN or offload them to the cloud for processing with a more complex and accurate CNN. This decision is made based on several factors such as network conditions, energy state of the mobile device, cloud backlog, and the hit-rate estimate. 
Ananthanarayanan \emph{et al.} \cite{Ananthanarayanan2019MobiSys} optimized their work based on the previous work \cite{Ananthanarayanan2017computer}. They executed a cheap CNN at the edge and a heavy CNN at the cloud. Only if the lightweight CNN model does not have sufficient confidence do they invoke the heavy CNN model.
Yi \emph{et al.}\cite{Yi2017SEC_LAVEA} presented LAVEA, a edge computing platform that utilizes serverless architecture to enable computation offloading between clients and edge nodes. They formulated the offloading task selection as an optimization problem to prioritize offloading requests in order to minimize response time.

\subsubsection{Partial Offloading}
Some researches partition the entire analytics task to each computing nodes, that is, the edge node shares part of the computing pressure for the cloud node to reduce end-to-end latency and resource consumption.
Kang \emph{et al.} \cite{Kang2017ASPLOS} proposed a system called Neurosurgeon to optimize the partitioning of deep neural network (DNN) computation between mobile and cloud. The system employs a series of models to predict the response time and power consumption of the model according to its configuration and type. This allows Neurosurgeon to lower the system latency, reduce mobile energy consumption, and enhance datacenter throughput.
Emmons \emph{et al.} \cite{Emmons2019HotEdgeVideo} proposed a concept of ``split-brain'' inference to perform video analytics. The approach involves processing the video partially on the camera, with the computation limited to a certain extent. Constrained by network capacity, the intermediate results are then transmitted to a cloud datacenter for further DNN inference.

\subsection{Collaborative Intelligence}\label{subsec_collab}

Computation offloading introduced above can be viewed as collaboration between nodes at different levels. We will then introduce collaborative intelligence, which refers to the collaboration between nodes at the same level.

\subsubsection{Mobile Collaboration}
Many districts are deploying cameras on a large scale, and when considering this scenario, we have to face the problem of increased massive deployment cost and effort. For example, when the overlapping angles of deployment of multiple cameras are large, or the same video analytics tasks are performed, a large amount of redundant data is generated and unnecessary computing resources are consumed.
In general, it is assumed that the computing power of the camera nodes is very limited, and it can only collect video streams, perform certain preprocessing and then send it to the edge or cloud. We have briefly introduced cross-camera collaboration in Sec \ref{subsubsec_fs} to optimize frame sampling and suppress redundancy \cite{Zhang2015MobiCom}. Next, we will introduce other research work on mobile collaboration.

Khochare \emph{et al.} \cite{Khochare2019CCGRID} developed Anveshak, a framework for creating video analytics applications that track objects across a camera network. In case an object is not detected by one camera, the system expands its search to include more cameras.
O'Gorman \emph{et al.} \cite{O'Gorman2018ICPR} quantified the temporal and spatial characteristics of video data and demonstrated how the sparse nature of real-world signals from public cameras can significantly reduce the computational load. This understanding can aid in making decisions to place the tasks on the edge or in the cloud.
Dao \emph{et al.} \cite{Dao2017ICDCS} proposed a framework that allows cameras coordination for high accuracy and reducing energy consumption. By coordination among cameras, each camera can use the non-optimal algorithm for the detection task and avoid unnecessary energy consumption.

Video analytics pipelines involve several adjustable parameters or ``knobs'', such as frame resolution, sampling rate, and detector models. The selection of these configurations has an impact on both the accuracy and the resources consumed of the video analytics. However, the number of potential configurations can increase exponentially with the number of knobs and their respective values. The Microsoft Research team is focusing on using cross-camera correlation to optimize the search space.
Jiang \emph{et al.} \cite{Jiang2018SIGCOMM_Chameleon} presented Chameleon, a controller that selects the optimal configuration for neural-based video analytics pipelines dynamically. The insight behind Chameleon is that the underlying characteristic that impacts the best configuration has enough spatio-temporal correlation, amortizing the search cost over time and across several video feeds.
Jain \emph{et al.} \cite{Jain2019HotMobile} proposed ReXCam, which leverages spatio-temporal correlations in video feeds from wide-area camera deployments to reduce search space of inference. By exploiting these correlations, ReXCam reduces both the workload and false positive rates in multi-camera video analytics. Specifically, ReXCam guides its search for a query identity by taking advantage of spatial and temporal locality dynamically of the camera networks.

\subsubsection{Edge Collaboration}
Compared with camera nodes, edge nodes have stronger computing power and can complete part of or even entire video analytics tasks. It is urgent to design task scheduling strategies or to share computation results among multiple edge nodes. Assigning a task to only one edge node may result in underutilization of the redundant computational resources of the other edge nodes. Hence, there is a need to explore optimal ways to group edge nodes and enable collaborative processing of video analytics tasks.

Long \emph{et al.} \cite{Long2017TMM} presented a framework for edge computing that enables cooperative processing of latency-sensitive multimedia tasks on resource-rich mobile devices. The framework relies on optimally dividing the mobile devices into different groups and assigning video chunks to the proper group to enable cooperative processing of tasks.
Wang \emph{et al.}\cite{Wang2017SmartIOT} designed a three-tiers architecture for real-time surveillance applications of edge computing system. The architecture is designed to elastically adjust computing capacity and dynamically route data to the most appropriate edge server. Surveillance tasks are executed by a group of Virtual Machines (VMs) or Virtualized Network Functions (VNFs) that work together on specific tasks. The system is effectively configured, monitored, and managed by the SDN controller.

\section{Resource management}\label{sec_resource}
Determination and detection of temporal and spatial events can be automatically done by video content analytics. This capability is supported by resources allocated to algorithms on cameras or devices to compute and output the object of interest \cite{Wang2018IC-NIDC, Lu18INFOCOM, Lu18TMC_CrowdVision}. Accordingly, three main dimensions are identified to characterize resource management in the context of video analytics, each consisting of various sub-dimensions. Hence, in a general overview, resource management is illustrated based on the following classification, and the summary is given in Table \ref{tb:rm_summary}.

\begin{table*}
\renewcommand\arraystretch{1.35}
	\caption{Summary of the literature for resource management}\label{CoCI}
	\centering
\linespread{1}\selectfont
		\begin{tabular}{|p{2cm}<{\centering}|p{2cm}<{\centering}|p{2cm}<{\centering}|p{2.5cm}<{\centering}|p{2cm}<{\centering}|p{2cm}<{\centering}|p{2cm}<{\centering}|}
			\hline \bf{Category} & \bf{Literature} & \bf{Method} & \bf{End Device Layer} & \bf{Edge/Fog Layer} & \bf{Cloud Layer} & \bf{Performance}\\
			\hline
			\hline
			\multirow{4}{*}{\shortstack{QoE Metrics}} & Hung \emph{et al.} \cite{ Hung2018SEC} & Narrowing down the configuration space & Choosing from configuration space for live video feed analytics & Determination of demand to the capacity ratio for configurations & \checkmark & Accuracy and response time \\
			\cline{2-7} &  Zhang \emph{et al.} \cite{Zhang2017nsdi} & Combination of offline profiler and online scheduler for resource-quality profile and allocation & Resource-quality profile of queries & $\times$ & \checkmark & Performance and lag \\
			\cline{2-7} & Lu \emph{et al.} \cite{Lu18INFOCOM,Lu18TMC_CrowdVision} & Batch processing and filtering out videos & Optimal video offloading and transmission sequence in terms of minimizing the query response time. & Deep learning techniques for object detection & $\times$ & Distributed processing, energy efficiency, and performance improvement \\
			\cline{2-7} & Lu \emph{et al.} \cite{Lu17MM} & Profiling and modeling resource requirements on the mobile platform for CNNs & Estimation of resource usage and compute time for CNN configurations  & $\times$ & $\times$ & Compute time \\
			\cline{2-7} & Lu \emph{et al.} \cite{Lu2020ToN_NetVision} & Usage of greedy algorithm and adaptive algorithm & Control transmission rate & Optimal video offloading and processing & $\times$ &Response time and transmission rate \\
			
			\cline{2-7} &Han \emph{et al.} \cite{Han2016MobiSys} & Heuristic scheduling algorithm & Frequency of use and the choosing the accurate model & $\times$ & DNN optimization at the cloud layer & Accuracy \\
			
			\cline{2-7} &Wang \emph{et al.} \cite{Wang2018IC-NIDC} & Light-weight edge cloud platform & $\times$ & Network switch and intelligent edge servers & Usage of container and SDN & Response time and QoE \\
			\cline{2-7} &Xu \emph{et al.} \cite{xu2018ATC} & Tuning canary inputs and usage of prediction  & Configuration selection considering video encoding parameters & $\times$ & $\times$ & Accuracy, low-latency and low-energy processing \\
			
			\cline{2-7} &Fu \emph{et al.} \cite{Fu2019ATC_EdgeWise} & Balancing operation queues by thread workers & Optimizing data flow by a congestion-aware scheduler & $\times$ & $\times$ & Latency \\
			\hline \multirow{4}{*}{\shortstack{Resource \\ Provisioning\\ Method}} & \cite{Yi2017SEC_LAVEA,Grassi2017SEC,Shen2017CVPR} &  Heuristic-based  & Localization algorithm or choosing among CNNs' models  & Usage of nearby devices for speeding up processing & $\times$ & Performance \\
			\cline{2-7} &\cite{Drolia2017ICDCS_Cachier,Long2017TMM,O'Gorman2018ICPR,Valls2018MILCOM} & Optimization-based & Load balancing  & Algorithm placement  & \checkmark  & Latency and performance\\
			\cline{2-7} &\cite{Huang18MM_QARC,Rao2017ICCV,Supancic2017ICCV} & DRL-based & Combination of deep reinforcement learning and ANN & $\times$ & $\times$ & Bitrate and network usage \\
			\cline{2-7} &\cite{Zhan2016MSN,Wu2017TMM,Hu2020TPDS} & Incentive-aware & Game theory strategies & $\times$ & $\times$ & Latency and cost \\
			\cline{2-7} & \cite{Stone2019SECON,Liu2018EdgeSys_EdgeEye,Poostchi2017CoRR} & CPU/GPU Optimization & Optimization techniques and libraries & $\times$ & $\times$ & Performance and cost \\
			\cline{2-7} &\cite{Xu2019arXiv_2,Huang2017sosp_SVE} & Data/Streaming Storage & Local storage & Maintaining cache servers & \checkmark & Latency and reducing network load \\
			\hline
\end{tabular}
\label{tb:rm_summary}
\end{table*}

\begin{itemize}
    \item \textit{Quality of Experience (QoE) Metrics} as a holistic concept is to measure a user satisfaction degree for a service or an application. This satisfaction can be fallen into three major groups: accuracy, latency, and performance (e.g., energy), taking into account edge or cloud resources. 
    \item \textit{Resources}  to \emph{be provisioned} for providing the guaranteed performance for application through selection, deployment, and runtime management of software and hardware resources contributing to video analytics.
    \item \textit{Resource provisioning methods} to \emph{deal} with providing the guaranteed performance for applications. This can be done by a heuristic-based resource provisioning scheme enforced by deep learning strategies, or it may consider content-aware approaches for resource allocation prioritization.  
\end{itemize}


\subsection{QoE Metrics}\label{sec:qoe}

\subsubsection{Accuracy}
Canary inputs (i.e., small input numbers, e.g., subsampling) have contributed to video accuracy optimization. Xu \emph{et al.} \cite{xu2018ATC} proposed VideoChef that uses canary inputs to tune the computation accuracy and transferring the approximate configurations to full inputs, which also uses a prediction model considering user constraints (e.g., Peak Signal-to-Noise Ratio) with respect to canary input's error. The canary inputs are chosen based on the dissimilarity ratio between the sample and the full-size video to determine how close a canary input is. This roots in the fact that most stages in the video pipeline are approximate, leading to low-latency and low-energy video processing in specific/generic domains. These approximations may depend on both algorithms and the video content, which necessitate exploring a large number of configurations for finding the optimal one. This selection takes place offline and through an efficient search strategy considering video encoding parameters. Since searching the best approximate level is \emph{computation-intensive}, VideoChef aims at an encoder-based approach emphasizing significant video changes through pixel and histogram-based for detecting the scenes.

Hung \emph{et al.} \cite{ Hung2018SEC} present VideoEdge as a system that is capable of achieving a trade-off between resource and accuracy through narrowing down the configuration space in a hierarchical structure (i.e., cameras, private clusters, and public clouds) for live video feed analytics. The configuration space is downsized by computation of maximum demand to the capacity ratio for each configuration to find the dominant demand, allowing the system to compare configurations across demand and accuracy. In addition, streaming and multi-programming can also lead to improving the accuracy as Han \emph{et al.} \cite{Han2016MobiSys} proposed a heuristic scheduling algorithm for allocating resources proportionally based on the frequency of use and selection of the most accurate model variant. 

Felemban \emph{et al.} \cite{Felemban2020TMC} proposed PicSys, a system that optimizes the deployment of a CNN pipeline by assigning each computation stage to resources that maximize their utilization. PicSys splits computation into several filtering stages, and assigns each stage to a resource using a heuristic algorithm that solves an optimization problem. To balance the trade-off between accuracy and speed, PicSys uses a lighter version of the CNN that requires less computation.

\subsubsection{Latency}
Wang \emph{et al.}, \cite{Wang2018IC-NIDC} proposed a lightweight edge computing platform that is powered by small-size and cost-efficient edge intelligent servers integrating computation and network capabilities to constitute a large scale edge cloud. The framework targets reducing response time while improving the quality of experiences. These goals are achieved by integrating software-defined network and container features. The latter provides network switches and the ease of services integration within containers for resource integration and controlling management.

The streaming processing engine presented in \cite{Fu2019ATC_EdgeWise} addressed improvement for latency through focusing on the operation queues to be balanced by thread workers. The engine that is called EdgeWise is powered by a congestion-aware scheduler that monitors the queues for selection of the highest operation priority to optimize the data flow. This optimization is mainly done by considering fixed worker pools for decoupling workers from operations (i.e., thread contention) stemming from ready threads to be in charge of operations with the most pending data. Moreover, EdgeWise considers data consumption policy for the queues to improve the overall scheduling decisions allocation of heavy-loaded operations with more frequent workers. The NetVision system \cite{Lu2020ToN_NetVision} used a greedy algorithm along with an adaptive algorithm to manage variate transmission rates to improve latency. It provided a solution for optimizing query response time in on-demand video processing scenario by formulating the processing scheduling problem.

\subsubsection{Performance}
Cloud resources may also contribute to performance improvement by pushing video processing to the cloud (i.e., offloading), which can be seen in \cite{Lu18INFOCOM,Lu18TMC_CrowdVision}. In these works, a deep learning-based CrowdVision platform is proposed that is distributed and energy-efficient and considers batch processing as the characteristics of CNNs. The features contribute to the distributed processing, balancing the waiting time of frame rates and the processing time of each batch for performance improvement while considering the network conditions for offloading benefits. To further improve the video processing, CrowdVision filters out videos by location and timestamp and applies deep learning techniques for the object of interest recognition in a batch-like manner. 
Zhang \emph{et al.} \cite{Zhang2017nsdi} proposed a video analytics system, VideoStorm, which leverages an offline profiler generating query resource-quality profile and employs an online scheduler to allocate resources for maximizing the performance on quality and lag rather than relying on fair sharing in clusters. The profiler uses greedy search and domain sampling specific for obtaining a set of handful configurations on the Pareto boundary of the profile, which the scheduler takes into account. The quality and lag are encoded as utility functions, which are penalized for violations and prioritization.

Artificial neural networks can also assist performance improvement. Lu \emph{et al.} \cite{Lu17MM} present Augur as a CNN performance analyzer to determine the efficiency of a CNN on a given mobile platform. This tool profiles and models the resource requirements of CNNs (i.e., the forward pass) by taking the configuration of CNN to estimate the computational overhead of the model.


\subsection{Resources}

\subsubsection{Hardware}
Power-efficient AI computing devices (i.e., AI-kit) have enabled video analytics at the edge and provide users with applications including image classification, object detection, segmentation, and speech processing running in parallel. These devices are also suitable for use in environments with intermittent connectivity, such as remote locations.

NVIDIA Jetson TX1 and TX2 \cite{jetsonnvidiatx2}, and Jetson Nano \cite{jetsonnvidianano} are power-efficient embedded AI computing devices that bring trustworthy AI computing at the edge. They have emerged from NVIDIA Pascal-family GPU capable of being integrated into any products due to having hardware standard features. 

Amazon Snowball \cite{awssnowball} is a device optimized for edge computing that provides virtual CPUs, block and object storage, and even an optional GPU. These devices can be clustered together in a rack-mounted form to create larger temporary installations. Snowball has been designed to support advanced machine learning and video analytics applications, especially in disconnected environments such as manufacturing, industrial, and transportation settings or in highly remote locations, including military or maritime operations.

Azure IoT Starter kit \cite{azureiot} is a vision AI developer kit to run AI models at intelligent edge. The kit runs models built by Microsoft Azure Machine Learning (AML) and other Azure services for edge analytics and AI processing.

Intel NUC mini PC \cite{intelnuc} is an AI Development Kitequipped with a powerful Intel Core processor, integrated graphics, and the advanced Intel Movidius Myriad X Vision Processing Unit (VPU). This combination allows for seamless execution of a wide range of AI workloads with high performance and low power consumption. The kit offers a comprehensive AI capability and is suitable for running diverse AI workloads.

Raspberry Pi \cite{raspi} is known for a series of small single-board computers that provides computing in a low cost, credit-card sized computer. There are also a majority of alternatives for this device which fall into single-board computers (SBCs) with powerful system-on-chip (SoC) such as Onion Omega2+ \cite{onionomega}, Orange Pi \cite{orangepi}, Banana Pi \cite{bananapi}, Rock64 \cite{rock}, Arduino \cite{arduino}, Asus Tinker Board \cite{tinkerboard}, Odroid \cite{odroid}, Pin64 \cite{pin64}, Cubieboard \cite{cubieboard}, BeagleBoard \cite{beagleboard}, LattePanda Alpha \cite{lattepanda}, UDOO BOLT \cite{udoobolt}, Libre Computer Le Potato \cite{librecomputer}, and NanoPi \cite{nanopi}.

\subsubsection{Software}

From a software perspective, there have been libraries for processing, such as OpenCV \cite{opencv} a cross-platform library that contains various image process and computer vision algorithm. This library also supports deep learning models (e.g., TensorFlow) and over 2500 optimized classic or state-of-the-art computer vision algorithms using MMX and SSE instructions when available. There are other frameworks alike OpenCV including SimpleCV \cite{simplecv} a Python-based open-source framework that is recommended for prototyping, and the other Python-based library is Scikit-image\cite{scikitimage} acting as a toolbox for SciPy and provides different algorithms for image processing. Accord.NET framework \cite{accordnetframework} that is a C\#-based framework providing several functionalities such as providing machine learning, computer vision, and image processing methods. BoofCV \cite{boofcv} as an open-source java library for real-time computer vision and robotics applications consisting of application-based packages for image processing, standard functions, and feature extraction algorithms. FastCV computer vision \cite{fastcv} that is implemented for ARM architecture and is optimized for Qualcomm’s Snapdragon processors for providing the most frequently used vision processing functions. MATLAB \cite{matlab} is a paid programming platform that comes with the computer vision processing toolbox. Deepface \cite{deepface} is a Python-based and lightweight face recognition and facial attribute analytics framework (e.g., age, gender, or emotion and race) that employs state-of-the-art models such as Google FaceNet \cite{schroff2015facenet} or VGG-Face \cite{vggface}. Point Cloud Library (PCL) is a C++-based library for three-dimensional image processing \cite{pcl}.  In addition, NVIDIA CUDA-X \cite{nvidialib1} provides libraries, tools, and technologies for delivering high-performance artificial intelligence application domains. There is also the NVIDIA Performance Primitives (NPP) \cite{nvidialib2} library that facilitates GPU-accelerated vision processing. Detectron2 \cite{detectron2} is Facebook artificial intelligence library for providing the latest technology and development in detection and segmentation algorithms, and supports computer vision research projects and production applications in Facebook.

Moreover, neural network frameworks (and related tools) have been introduced to the image processing which have been used widely such as YOLOv3 \cite{yolov3}, MobileNet \cite{howard2017mobilenets}, ResNet \cite{he2016identity}, VGG \cite{simonyan2014very}, NASNet \cite{zoph2018learning}, PNASNet \cite{liu2018progressive}, Keras \cite{kerasapi}, Caffe \cite{cafeapi}, PyTorch \cite{pytorch}, Albumentations \cite{albumentations}, OpenVINO \cite{openvino}, and TensorFlow \cite{tensorflow}. These are based on the convolutional deep neural network for the purpose of object of interest detection.

\subsection{Resource Provisioning Methods}

\subsubsection{Heuristic-based}
Previous studies mentioned in Section \ref{sec:qoe} utilize heuristic methods to improve the QoE, e.g., \cite{Lu2020ToN_NetVision, Felemban2020TMC, Hung2018SEC, Zhang2017nsdi, Hung2018SEC, Han2016MobiSys}. There still exist other research studies whose aims lead to better resource allocation. 

Grassi \emph{et al.} \cite{Grassi2017SEC} proposed a method to estimate the location of parked cars in a single frame using a localization algorithm. The approach involves camera calibration against well-known objects in the surrounding environment.  In contrast, Yi \emph{et al.} \cite{Yi2017SEC_LAVEA} proposed a light-weight virtualization on top of the operating system. It is a two-phase optimization process in which, in the first phase, bandwidth is allocated, and the next phase aims to leverage nearby edge resources to expedite the task completion time. Shen \emph{et al.} \cite{Shen2017CVPR} leveraged short-term class skew (i.e., objects of interests over some time) to accelerate video classification using CNNs. The research proposes a heuristic algorithm based on exploration and exploitation strategies to address the sequential model selection problem, which is formulated as the Oracle Bandit Problem (OBP). The proposed approach seeks to estimate skew at test-time, producing a specialized model if possible and using the model as long as the skew lasts and then reverts to one of the classifier models referred to as oracle (e.g., GoogleNet). To create a specialized model, the research study selects a specified number of dominant classes and a randomly selected subset from other classes with a different label from the original data, creating new training data sets. The dominant classes make up a percentage of the new dataset used to train the compact model. The oracle model is swapped with a less expensive but compact model for exploiting skew to return early with the classification result if inputs belong to the frequent classes in the incoming distribution. Otherwise, it uses the oracle model.

The research proposes a heuristic algorithm based on exploration and exploitation strategies to address the sequential model selection problem, which is formulated as the Oracle Bandit Problem (OBP). 

\subsubsection{Optimization theory-based}

Similar to \cite{Lu18INFOCOM, Lu18TMC_CrowdVision,Fu2019ATC_EdgeWise,Yi2017SEC_LAVEA}, 
Drolia \emph{et al.} \cite{Drolia2017ICDCS_Cachier} applied tunable cache size as the knob to minimize the latency. The authors presented Cachier, a system that optimizes the cache size to minimize latency in computation-intensive recognition applications. The authors model edge servers as caches and use novel optimizations to adaptively balance load between the edge and the cloud.
Long \emph{et al.} \cite{Long2017TMM} proposed a solution for the group formation problem by transforming it into a winner determination problem. This new formulation can be solved using a 2-approximation algorithm that significantly reduces the complexity of the problem.

O'Gorman \emph{et al.} \cite{O'Gorman2018ICPR} studied the upper-bound requirements for processing and bandwidth for a video analytics application analytically and experimentally on real videos to provide guidance for algorithm placement between the edge and the cloud. However, Valls \emph{et al.} \cite{Valls2018MILCOM} modeled the resource allocation as a network processing model in which targets the order of processing, where two resource allocation algorithms are presented due to the dynamicity of the network. The proposed approach utilizes two algorithms: a backpressure-based algorithm and a backpressure plus interior-point method algorithm. The former allocates data analytics resources based solely on the congestion level in the system, while the latter considers the fact that some resources have a constant demand.

\subsubsection{DRL-based}
Huang \emph{et al.} \cite{Huang18MM_QARC} proposed a rate deep reinforcement learning algorithm focusing on a higher perceptual quality rate with lower bitrate. Based on a trained neural network, the algorithm predicts future bitrates considering the observation of current network status. The model is divided into two separate neural networks; an ANN to predict future video quality based on the previous video frame and the other ANN as a reinforcement learning one that determines the bitrate with the first model's result. The first model has a combination of CNNs to extract image features and a recurrent neural network for capturing temporary features. 
Rao \emph{et al.} \cite{Rao2017ICCV} introduced a deep reinforcement learning method for video face recognition that leverages attention mechanisms. They approached the problem of identifying relevant regions of a video as a Markov decision process and trained an attention model through a DRL framework, which does not require additional labeling.
Supancic \emph{et al.} \cite{Supancic2017ICCV} modeled online video object tracking and transformed the tracking formula into a partially observable decision-making process with DRL to understand the best decision-making strategy.

\subsubsection{Incentive-aware}
Zhan \emph{et al.} \cite{Zhan2016MSN} presented a scheme for optimizing the interaction between video providers and mobile users, by formulating it as a two-person cooperative game. The proposed scheme used Nash bargain game theory to obtain the optimal cooperation decision, resulting in a high data delivery ratio, low communication and computation overhead, and excellent economic properties.
Wu \emph{et al.} \cite{Wu2017TMM} proposed a pricing mechanism based on Stackelberg game theory to inspire device-to-device video distribution from core users. This approach can reduce the load on base stations and improve the effectiveness and reliability of video transmission.
Hu \emph{et al.} \cite{Hu2020TPDS} present a solution to the challenge of offloading heterogeneous video analytics tasks that leverages game theory. They cast the problem as a minority game, in which each participant must make decisions independently in each round, and the players who make the minority choice win. The game is played by multiple players who lack complete information sucn as the number of video analytics tasks or resources, which creates incentives for players to cooperate with each other.

\subsection{Other Optimizations}
In the following, some other optimization studies in video analytics not mentioned above will be introduced, including GPU/CPU acceleration and data/streaming storage.

\subsubsection{GPU/CPU acceleration}
To reduce the costs of supporting a large-scale deployment with hundreds of video cameras, Stone \emph{et al.} \cite{Stone2019SECON} introduced Tetris, a system that incorporates various optimization techniques from computer vision and deep learning fields in a synergistic manner.
Liu \emph{et al.} \cite{Liu2018EdgeSys_EdgeEye} simplify the memory transfer between CPU and GPU with Nvidia CUDA mapped memory feature. The DetectNet component leverages TensorRT to manage the GPU inference engine, including tasks such as initializing the inference engine, creating the inference runtime, loading the serialized model, creating the inference execution context, and executing the inference engine.

Integral histograms are widely used for extracting multi-scale histogram-based regional descriptors, which are essential components of many video content analytics frameworks. Poostchi \emph{et al.} \cite{Poostchi2017CoRR} evaluated different approaches to mapping the computation of integral histograms onto GPUs, using various kernel optimization strategies.

\subsubsection{Data/Streaming Storage}

Maintaining a cache server in the edge node can directly provide video service-related content to users and at the same time, reduce the number of requests forwarded to back-end servers. It can
minimize the expected latency for users and reduce network load and back-end load. The storage system can store various data ingested and generated during the video encoding process as this data can play a role in the subsequent analytics, and the video service can be pushed to the user faster.
Huang \emph{et al.} \cite{Huang2017sosp_SVE} presented a streaming video engine that uses a preprocessor to store original upload videos in parallel onto disk for fault tolerance. To process the video data, the engine employs a scheduler that schedules tasks on workers. These workers pull data from the preprocessor and store the processed data onto disk.

The term ``zero-streaming" camera refers to cameras that capture videos to local storage without streaming anything. They are reactive and highly efficient, consuming only network and cloud resources to analyze queried video footage.
Xu \emph{et al.} \cite{Xu2019arXiv_2} exploited the zero-streaming paradigm and minimized the ingestion cost, shifting as much as possible to the query execution. Zero-streaming cameras capture videos to the local flash storage (cheap and large) without uploading any; only in response to user queries, they communicate and cooperate with the cloud to analyze the stored videos.

\section{Security and Privacy}\label{sec_privacy}
\begin{table*}
\renewcommand\arraystretch{1.35}
	\caption{Summary of the security and privacy methods}\label{privacy}
	\centering
\linespread{1}\selectfont
		\begin{tabular}{|p{2.5cm}<{\centering}|p{2cm}|p{2.5cm}|p{3cm}<{\centering}|p{3cm}<{\centering}|p{2.5cm}<{\centering}|}
			\hline \bf{Category} & \bf{Literature} & \bf{Method} & \bf{End Device Layer} & \bf{Edge/Fog Layer} & \bf{Cloud Layer} \\ \hline
			\hline \multirow{6}{*}{\shortstack{Privacy-preserving\\ Video Collection}} & Kim \emph{et al.} \cite{Kim2014ICUAS} & Filtering, Encryption & Filtering frames according to preset privacy map & Applying encrypt operation& Applying decrypt operation \\
			\cline{2-6} &  \cite{Wang2017MMSys, Wang2018TOMCCAP} & Denaturing, Advanced encryption system (AES) & $ \times  $ & Using denaturing to blur faces and encrypt raw video & Data decryption \\
			\cline{2-6} & Wang \emph{et al.} \cite{Wang2019Access} &  Advanced encryption system (AES)  & $\times $ & Using AES to generate secret face feature & Split the secret feature into all edge servers for storing  \\
			\cline{2-6} & Li \emph{et al.} \cite{Li2018IOTJ} & Encryption & Encrypting raw video & Verifying and aggregating the encrypted data & Data decryption\\
			\cline{2-6} & Zarepour \emph{et al.} \cite{Zarepour16PerCom} & Filtering & Applying filtering operation & $\times$ & $\times$ \\
			\cline{2-6} & \cite{wang2019arXiv, xu2018Communications_Magazine,mao2018USENIX} & Differential Privacy & Adding differential noise & Analyze obfuscated data & Analyzing obfuscated data \\
			
			\hline \multirow{5}{*}{\shortstack{Privacy-preserving \\Video Analytics}} & \cite{Jiang17DSC,Khazbak2019IoTDI,Li2019FCS} & Homomorphic Encryption  & $\times$ & Homomorphic Encryption & Video analytics on encrypted data directly \\
			\cline{2-6} & Akkaya \emph{et al.} \cite{akkaya2019CNS} & Background Subtraction, Homomorphic Encryption & Background subtraction on UAV & Homomorphic encryption & Further analytics on encrypted data \\
			\cline{2-6} & \cite{sada2019DASC, chen2019GLOBECOM, liu2020arXiv} & Federated Learning & $\times$ & The federated learning layer deployed on edge & $\times$ \\
			\cline{2-6} & \cite{xu2019arXiv_3, osia2020IOT} & NN-based Obfuscation & $\times$ & Using NN to extract features from video frames & Further analytics on frame features \\
			\cline{2-6} & Hu \emph{et al.} \cite{Hu2021Springer} & DP-empower Federated Learning & $\times$ & Training local model at the edge and conduct perturbation on model parameters & maintaining public model  \\

			\hline \multirow{2}{*}{\shortstack{Privacy-preserving \\Video Storage}} & \cite{Nigel2016HotMobile,Wang2017MMSys} &  Encryption  & $\times$ & Trusted edge servers provide data storage & $\times$ \\
			\cline{2-6} & \cite{Wang2020JPDC, Xiao2020Sensors} & Frame data dividing& $\times$ & Dividing video data into different parts and storing on edge layer &  $\times$ \\
			\hline
	\end{tabular}
\end{table*}

People are extremely concerned about privacy leakage, especially for video applications with rich personal information. Privacy protection enables individuals to have a certain degree of control over their sensitive data, preventing it from being abused by third parties. In cloud-based video analytics, the privacy-preserving operations need to be locally executed, which demands a high level of computation resources and a low complexity on the encryption algorithms. With the help of edge servers, the privacy-preserving operations can be executed at edge servers, avoiding straight exposure of privacy information at the cloud. For a video analytics-based application, massively captured video data commonly go through the following three phases: \emph{Data Collection}, \emph{Data Analytics} and \emph{Data Storage}. Different phases suffer from different privacy risks and corresponding privacy preserving mechanisms are required for privacy protection. In this section, we summarize the related works on these three categories.

\subsection{Privacy-preserving Video Collection}

In the stage of data collection, privacy concerns can come from unreliable application owners as well as unreliable network conditions. For video analytics applications, some sensitive information (e.g., facial information and the ID number) should not be collected by an untrusted party in avoid of being illegally used. Besides, in over-the-air transmission of raw video contents, there is a giant risk of information leakage against eavesdroppers. Considering these issues, it is essential to design privacy-preserving video collection solutions. As an efficient and proven solution, it is the features not the raw contents that should be transmitted to the edge or cloud servers. For example, it is sufficient to transmit the outline of a pedestrian without the need for facial information in a pedestrian counting application. Thus, the facial information as the private source should not be exposed to the central cloud server, and thus, the privacy can be reserved. Based on the integrity of the collected video contents, we summarize recent works as \emph{complete collection} and \emph{partial collection} as follows.

\subsubsection{Complete Collection}

Encryption methods are widely used for a full collection of video contents.
Kim \emph{et al.} \cite{Kim2014ICUAS} encrypted video frames on edge servers in the surveillance station and then delivered encrypted frames to a trusted third-party cloud server.  The specific regions of a video frame are first filtered out according to a preset privacy map and then decrypted into original video data with a shared key at the cloud server. However, privacy map cannot provide a more fine-grained and automated privacy protection. Besides, it does not always  provide a trusted third-party server as a medium between video data and video subscribers. 
Li \emph{et al.} \cite{Li2018IOTJ} proposed a privacy-preserving data aggregation scheme, where the edge server aggregates the encrypted data from terminal devices and sends data to the cloud server, which can then decrypt the aggregated data through its private key. In this kind of privacy-preserving systems, edge servers always serve as an intermediate layer to provide privacy protection for the entire system. 

Traditional encryption methods are generally computation-intensive, which might not work well on resource-limited edge servers. 
To address such deficiency, Wang \emph{et al.} \cite{wang2019arXiv} proposed a VideoDP platform that provides a novel differential privacy (DP) function. In VideoDP, adding or removing any sensitive visual element into/from the input video does not significantly affect the analytical result.
Xu \emph{et al.} \cite{xu2018Communications_Magazine} proposed a local DP obfuscation framework for data analytics, where data is distilled in edge servers with limited ability to make inferences about users' sensitive data. 
Mao \emph{et al.} \cite{mao2018USENIX} proposed to partition a DNN model after the first convolutional layer between the end device side and the edge server side. Then, DP is applied to protect  convolutional layers to guarantee the privacy of users' sensitive data.

\subsubsection{Partial Collection}

Recall the frame cropping technology introduced in Sec. \ref{sec_frame_crop}, many video analytics-related applications focused on RoIs in each video frame, such as the face regions in face detection and the vehicle regions in vehicle monitoring. Thus,  end devices (e.g., cameras) should also enhance the privacy preservation in partial video collection.

Neural networks can offer more granular privacy protection measures. OpenFace \cite{Wang2017MMSys, Wang2018TOMCCAP} implemented a privacy-preserving data collection mechanism by denaturing video data on the edge server instead of directly sending the raw video to the cloud. This technology selectively blurs faces that appear in video frames to alleviate privacy concerns related to face data. Similarly, Wang \emph{et al.} \cite{Wang2019Access} proposed a privacy-preserving face verification system, which applies the edge server to extract face feature by CNNs, as well as encryption of the feature data before sending to the cloud server by using advanced encryption system (AES). Zarepour \emph{et al.} \cite{Zarepour16PerCom} introduced a novel context-aware privacy-preserving framework, which uses the contextual information to estimate the set of potential sensitive subjects in each image. In this framework, user activity extraction and sensitive information filtering are completed on the end device locally before publishing the raw image.

\subsection{Privacy-preserving Video Analytics}

During video analytics, edge servers with limited computing power may need to process sensitive information on untrusted platforms. Therefore, before the original data is submitted to a untrusted platform for further analysis, corresponding preprocessing should be performed at the edge, such as encryption or abstract knowledge extraction. A crucial question is how to apply the computing power of the cloud server without exposing the privacy of raw video. There are three technologies that can be possible solutions, which are \emph{encryption-based technology}, \emph{obfuscation-based technology} and \emph{privacy-preserving machine learning}.

\subsubsection{Privacy-preserving Training}

In an edge-based video analytics system, data can be easily collected by terminal devices. However, plain-text data cannot be directly sent to the cloud server when users are concerned about privacy. Instead, an edge server can use its own data to conduct model training locally without sharing data with the cloud server, or use a neural network to perform preprocessing.

The use of edge computing has enabled video analytics applications to benefit from low-latency and distributed data processing services by leveraging the storage and computing capabilities of nearby end devices. However, in a distributed environment, the training process remains challenging due to the fragmented knowledge base across edge and cloud servers.
As a promising solution, \emph{federated learning} enables a distributed server to train its model locally and does not need to share local data with others. Thus, federated learning greatly alleviates the risk of privacy leakage caused in data sharing. 
Sada \emph{et al.} \cite{sada2019DASC} proposed an edge-based video analytics architecture, using federated learning to update object detection models and avoid sending the local data to the cloud. The federated learning layer is deployed in an edge server, which is situated between end devices and the cloud server. 
Similarly, Chen \emph{et al.} \cite{chen2019GLOBECOM} proposed a distributed learning framework that can be trained at each base station and cooperatively builds a learning model which can predict the mobility and orientations of users. 
Liu \emph{et al.} \cite{liu2020arXiv} developed a platform called FedVision, which allows for the development of federated learning powered computer vision applications. It aims to develop effective visual object detection models by utilizing image data owned by multiple organizations through federated learning. FedVision is the first industry application of FL in computer vision-based tasks, and it has the potential to assist organizations in complying with stricter data privacy protection laws, such as GDPR.

Although federated learning does not require the transmission of raw data during training, the attacker may still obtain user privacy from the exposure of gradients. For example, \cite{melis2019sp} shows that an attacker can infer whether a participant's data has been included in the dataset by collecting and analyzing shared models with an accuracy of 90\%. 
In federated learning paradigm, there are several methods that can be used to improve the video analytics privacy-preserving level. The main methods include the \emph{homomorphic encryption} method and the \emph{differential privacy} method.

\emph{(i) Homomorphic encryption-empowered solution}

Homomorphic encryption \cite{FHE_Gentry2009} enables data to be analyzed and manipulated while still encrypted, without the need for decryption.
Jiang \emph{et al.} \cite{Jiang17DSC} explored the use of homomorphic encryption to perform scale-invariant feature transform on encrypted images, enabling data analytics to be performed directly on encrypted data. To enable homomorphic encryption operations on resource-constrained edge servers, TargetFinder \cite{Khazbak2019IoTDI} applied optimization techniques to reduce computation overhead of cryptography primitives. These technologies enable secure and privacy-preserving image processing and analysis on edge devices with limited resources.
Li \emph{et al.} \cite{Li2019FCS} proposed a novel framework for privacy-preserving computing that utilizes lightweight permutation-substitution encryption and homomorphic encryption on end devices. This edge-assisted framework offloads the burden of computation, communication and storage while ensuring data security.
Akkaya \emph{et al.} \cite{akkaya2019CNS} proposed to perform background subtraction to get the foreground to transmit, and thus reduce the size of the data to be transmitted and the computational cost for applying homomorphic encryption. In this way, the receiver can aggregate the background and foreground to do further analytics solely based on the encrypted data.
Ma \emph{et al.} \cite{ma2018INFOCOM} presented a privacy-preserving motion detection algorithm for HEVC compressed videos which operates in the compressed domain. It can detect the coarse-grained shapes of moving objects and estimate the motion trajectory without decoding the video. By searching in the compressed-domain, the algorithm can preserving the compression efficiency of the video codec without incurring extra transmission bandwidth or storage overhead.

\emph{(ii) Differential privacy-empowered solution}

Without the high computation burden of the homomorphic operation, differential privacy provides a lightweight solution for the federated learning paradigm. Differential privacy-empowered solution adds zero-mean ``noises" to the trained parameters by using some randomized mechanism which is called differential privacy-preserving, such as Gaussian mechanism. 
Hu \emph{et al.} \cite{Hu2021Springer} proposed \emph{FedEVA}, a distributed training framework for edge video analytics that protects user privacy with fast convergence rates. The framework implements local differential privacy (LDP) on user updates before sending  gradients to the parameter server, which then updates the neural network model based on the perturbed gradients. Experimental results shows that the proposed framework can ensure privacy preservation while maintaining the same convergence rate.

\subsubsection{Privacy-preserving inference}

We have seen that model partition technologies can extract abstract features of raw video data on the edge layer, while further analytics tasks are completed on the cloud. However, input recovery attacks can occur during inference, aiming to recover raw image data from image features. \emph{Privacy-preserving inference} pays more attention to resisting input recovery attacks during the model inference phase. 
Chi \emph{et al.} \cite{chi2018arXiv} proposed a framework to ensures user data privacy during the model inference stage, when users utilize their data to obtain classification results. This framework addresses the potential for unauthorized access to privacy-sensitive information like encoding information about previous layers, which may occur due to the presence of multiple intermediate layers in the output of a deep learning neural network. 
Osia \emph{et al.} \cite{osia2017privacy} introduced a method to manipulate the extracted features by altering the training phase when applying the Siamese network \cite{chopra2005learning} and a noise addition mechanism for improving privacy protection. Moreover, they applied transfer learning and deep visualization techniques to quantify the privacy guarantees of their approach. 
Similarly, Osia \emph{et al.} \cite{osia2020IOT} proposed a novel approach that leverages an edge device to run the initial layers of a neural network for protecting user privacy. The output is then sent to the cloud to process the remaining layers and produce the final results. To further improve privacy protection, they employed Siamese \cite{chopra2005learning} fine-tuning to ensure that only necessary information is contained on the user's device for the main task, thus preventing any secondary inference on the data.
Xu \emph{et al.} \cite{xu2019arXiv_3} presented a lightweight and unobtrusive approach to obfuscating the inference data at user devices. The edge servers only need to execute a lightweight neural network to obfuscate the inference data implying that thus the neural network can be easily deployed on a resource constrained edge server or device introducing light compute overhead.

\subsection{Privacy-preserving Video Storage}

Reliable data storage is important for video analytics tasks, especially for video retrieval applications. Privacy concerns come from sending sensitive data directly to the cloud with a lack of user control. As we know, more and more end devices have the ability to collect high-definition video content with a large data size, which inevitably causes the storage problem. Unlike mobile devices, edge and cloud servers have more powerful  storage and computing capabilities. However, the pure video storage on the remote servers will cause the leakage of  data privacy.
Davies \emph{et al.} \cite{Nigel2016HotMobile} proposed a software solution called Privacy Mediator, which is in the same administrative domain with  end devices. Therefore, Privacy Mediator can provide a reliable data storage service. OpenFace\cite{Wang2017MMSys} ensures a stronger privacy protection with a similar way by applying a trusted edge server to perform video denaturing and provide edge-based data storage, which keeps  privacy data away from unreliable network transmission. 

To protect privacy, Neff \emph{et al.} \cite{neff2019IOT} proposed REVAMP2T, which does not store or transfer any image data across the network. The edege server will destroy the image as soon as the image is processed. Instead, it works on an encoded feature representation of an individual, which has no meaning outside of the REVAMP2T system and cannot be interpreted by humans. However, it is important to utilize the computing power on the cloud side when the task suffers from a high computational complexity. 
Wang \emph{et al.} \cite{Wang2020JPDC} proposed a three-layer storage architecture, in which edge servers can offer a computing and storage service while the rest of data is transmitted to the cloud. With the proposed storage architecture, privacy data cannot be retrieved even when we use a cloud storage service. 
Similarly, Xiao \emph{et al.} \cite{Xiao2020Sensors} fully utilized the storage space of edge and cloud servers by proposing a hierarchical edge computing architecture and they divide video frames into three parts. To be specific, the most significant bits of key frames are stored in local with full control and the least significant bits are encrypted before sending to the edge servers. Finally, non-key frames are compressed and encrypted before they are transmitted to the cloud. For one thing, the above architecture makes full use of different levels of storage space; for another, it also provides a more fine-grained level of privacy protection.

\section{Issues and Future Research Directions}\label{sec_issue}
This section presents the essential issues and future research directions in the area of edge-based video analytics.

\subsection{Efficient Video Compression Methods}
In the design of video transmission, it has been proved that an efficient video compression can tremendously reduce the bandwidth consumption.
However, it has not been effectively revealed how to describe the relationship between the video transmission efficiency and the video analytics accuracy.
For example, when the video is coded via the super resolution technology that can be transmitted without consuming much bandwidth, how we can quantitatively obtain the video analytics performance, e.g., accuracy.

\subsection{5G/6G Empowered Video Analytics}
In the video analytics applications, the focused parts are always the mobile objects on the captured frames. However, it has not been given enough attention on the use cases that the cameras themselves are in motion. Although some preliminary works focused on the scenarios that cameras are installed on the drones, challenges still remain, especially for the emerging 6G technologies that support satellite services.

In the future 6G service scenario, the video analytics applications will not only process the video streaming captured by the cameras deployed at stores, crossroads, and places all around the cities, but also by the cameras installed on the satellite. 
Denby \emph{et al.} \cite{Denby2020ASPLOS_OEC} proposed an orbital edge computing system to enable on-board edge computing at each nano-satellite with camera, allowing for local processing of sensed data when downlinking is not feasible. It is urgently needed to take as a coupled consideration with all orbit parameters, physical models, and ground station positions to trigger data collection, predict energy availability, and conduct task offloading, and execute video analytics tasks.

Current works generally focused on the video analytics use cases and service scenarios for the civil use. Most researchers developed the systems by balancing the trade off between accuracy and latency. However, some dedicated industries and applications put high requirements for accuracy and latency, e.g., the video analytics of track safety monitoring for high-speed railway. Thus, it is urgently required to design a dedicated edge-based video analytics architecture for these kinds of applications.

\subsection{Interactive Video Analytics System}
Augmented reality technology allows for interactive elements to be added over real-world views for specific purposes. However, AR overlays digital is overlaid onto the real world views and the digital content is not directly anchored to real-world elements, which means that it cannot interact with real-world elements. To overcome this limitation, AR researchers must develop techniques that allow digital content to interpret and respond to users' head movements and body gestures dynamically, enabling a more interactive and immersive AR experience.

\section{Conclusion}
Edge-based video analytics have emerged as one of the most widely adapted implementations for various smart services, entertainment, safety and security. This survey has made an overarching taxonomy that covers key aspects of edge-based video analytics with respect to use cases, architectures, techniques, resource management, and security and privacy. Besides, some recommendations on future research issues and directions are provided. In particular, we have identified edge-based solutions have the following main advantages. 
\emph{Firstly}, a low response time is urgently required for most video analytics applications. However, local execution (executed on end devices) efficiency is harsh restricted by the computation resources. Differently, the tasks executed on the cloud are rarely affected by the computation resources. Nevertheless, the video content transmission requires high bandwidth, which is hard to achieve by the cloud computing. Edge computing brings computation and communication closer to the task source, and improves response times.
\emph{Secondly}, located between end devices and remote cloud, edge servers provide a new dimension of dynamical resource allocation that can promote system optimization.
The resource provisioning scalability from edge servers provides greater potential to achieve better service performance.
\emph{Thirdly}, edge servers can provide more security and privacy guarantee. Due to the storage limitation, the video contents cannot be all locally stored, and the edge servers can be a good candidate providing privacy-preserving storage solutions. Compared to the video collection, analytics, storage at the cloud, the edge-based solution can provision better privacy preservation due to its distributed properties.
As a result, edge-based video analytics solutions can be leveraged as a general framework to support several privacy-preserving video analytics applications. 

\balance

\Urlmuskip=0mu plus 1mu\relax
\bibliographystyle{IEEEtran}
\bibliography{EVA}

\begin{thebibliography}{100}
\providecommand{\url}[1]{#1}
\csname url@samestyle\endcsname
\providecommand{\newblock}{\relax}
\providecommand{\bibinfo}[2]{#2}
\providecommand{\BIBentrySTDinterwordspacing}{\spaceskip=0pt\relax}
\providecommand{\BIBentryALTinterwordstretchfactor}{4}
\providecommand{\BIBentryALTinterwordspacing}{\spaceskip=\fontdimen2\font plus
\BIBentryALTinterwordstretchfactor\fontdimen3\font minus
  \fontdimen4\font\relax}
\providecommand{\BIBforeignlanguage}[2]{{%
\expandafter\ifx\csname l@#1\endcsname\relax
\typeout{** WARNING: IEEEtran.bst: No hyphenation pattern has been}%
\typeout{** loaded for the language `#1'. Using the pattern for}%
\typeout{** the default language instead.}%
\else
\language=\csname l@#1\endcsname
\fi
#2}}
\providecommand{\BIBdecl}{\relax}
\BIBdecl

\bibitem{ihs2015}
\BIBentryALTinterwordspacing
``Video surveillance: New installed based methodology yields revealing
  results,'' 2015. [Online]. Available:
  \url{https://ihsmarkit.com/pdf/IHS-Video-surveillance-installed-base(2)_227038110913052132.pdf}
\BIBentrySTDinterwordspacing

\bibitem{fortune}
``Video analytics market analysis.''
  \url{https://www.fortunebusinessinsights.com/industry-reports/video-analytics-market-101114}.

\bibitem{Yi2017SEC_LAVEA}
S.~Yi, Z.~Hao, Q.~Zhang, Q.~Zhang, W.~Shi, and Q.~Li, ``{LAVEA}: Latency-aware
  video analytics on edge computing platform,'' in \emph{Proceedings of IEEE
  37th International Conference on Distributed Computing Systems (ICDCS)},
  2017, pp. 2573--2574.

\bibitem{Hung2018SEC}
C.-C. Hung, G.~Ananthanarayanan, P.~Bodik, L.~Golubchik, M.~Yu, P.~Bahl, and
  M.~Philipose, ``Videoedge: Processing camera streams using hierarchical
  clusters,'' in \emph{Proc. of IEEE/ACM Symposium on Edge Computing (SEC)},
  2018, pp. 115--131.

\bibitem{Shi2016IEEEInternet}
W.~Shi, J.~Cao, Q.~Zhang, Y.~Li, and L.~Xu, ``Edge computing: Vision and
  challenges,'' \emph{IEEE Internet of Things Journal}, vol.~3, no.~5, pp.
  637--646, 2016.

\bibitem{Zhou2019IEEE}
Z.~Zhou, X.~Chen, E.~Li, L.~Zeng, K.~Luo, and J.~Zhang, ``Edge intelligence:
  Paving the last mile of artificial intelligence with edge computing,''
  \emph{Proceedings of the IEEE}, vol. 107, no.~8, pp. 1738--1762, 2019.

\bibitem{Vega2018IEEE}
M.~T. Vega, C.~Perra, F.~De~Turck, and A.~Liotta, ``A review of predictive
  quality of experience management in video streaming services,'' \emph{IEEE
  Transactions on Broadcasting}, vol.~64, no.~2, pp. 432--445, 2018.

\bibitem{Barakabitze2019IEEE}
A.~A. Barakabitze, N.~Barman, A.~Ahmad, S.~Zadtootaghaj, L.~Sun, M.~G. Martini,
  and L.~Atzori, ``{QoE} management of multimedia streaming services in future
  networks: a tutorial and survey,'' \emph{IEEE Communications Surveys \&
  Tutorials}, vol.~22, no.~1, pp. 526--565, 2019.

\bibitem{Jedari2020surveys}
B.~Jedari, G.~Premsankar, G.~Illahi, M.~Di~Francesco, A.~Mehrabi, and
  A.~Yl{\"a}-J{\"a}{\"a}ski, ``Video caching, analytics and delivery at the
  wireless edge: A survey and future directions,'' \emph{IEEE Communications
  Surveys \& Tutorials}, 2020.

\bibitem{Jun2017SJ}
S.~{Jun}, T.~{Chang}, H.~{Jeong}, and S.~{Lee}, ``Camera placement in smart
  cities for maximizing weighted coverage with budget limit,'' \emph{IEEE
  Sensors Journal}, vol.~17, no.~23, pp. 7694--7703, 2017.

\bibitem{Hu2018IOTJ}
L.~{Hu} and Q.~{Ni}, ``{IoT}-driven automated object detection algorithm for
  urban surveillance systems in smart cities,'' \emph{IEEE Internet of Things
  Journal}, vol.~5, no.~2, pp. 747--754, 2018.

\bibitem{Zhang2017nsdi}
H.~Zhang, G.~Ananthanarayanan, P.~Bodik, M.~Philipose, P.~Bahl, and M.~J.
  Freedman, ``Live video analytics at scale with approximation and
  delay-tolerance,'' in \emph{Proc. of 14th USENIX Symposium on Networked
  Systems Design and Implementation (NSDI)}, 2017, pp. 377--392.

\bibitem{Ananthanarayanan2019MobiSys}
G.~Ananthanarayanan, V.~Bahl, L.~Cox, A.~Crown, S.~Nogbahi, and Y.~Shu, ``Demo:
  Video analytics-killer app for edge computing,'' in \emph{ACM MobiSys}, 2019.

\bibitem{Grassi2017SEC}
G.~Grassi, K.~Jamieson, P.~Bahl, and G.~Pau, ``Parkmaster: An in-vehicle,
  edge-based video analytics service for detecting open parking spaces in urban
  environments,'' in \emph{Proceedings of the Second ACM/IEEE Symposium on Edge
  Computing (SEC)}, 2017, p.~16.

\bibitem{Xie2018CyberC}
Y.~Xie, Y.~Hu, Y.~Chen, Y.~Liu, and G.~Shou, ``A video analytics-based
  intelligent indoor positioning system using edge computing for iot,'' in
  \emph{Proc. of International Conference on Cyber-Enabled Distributed
  Computing and Knowledge Discovery (CyberC)}, 2018, pp. 118--1187.

\bibitem{Barthelemy2019Sensors}
J.~Barth{\'e}lemy, N.~Verstaevel, H.~Forehead, and P.~Perez, ``Edge-computing
  video analytics for real-time traffic monitoring in a smart city,''
  \emph{Sensors}, vol.~19, no.~9, p. 2048, 2019.

\bibitem{smartfarm}
\BIBentryALTinterwordspacing
\emph{Precision agriculture and the future of farming in Europe Scientific
  Foresight Study}, 2016. [Online]. Available:
  \url{https://www.europarl.europa.eu/RegData/etudes/STUD/2016/581892/EPRS_STU(2016)581892_EN.pdf}
\BIBentrySTDinterwordspacing

\bibitem{Alharbi2021Access}
H.~A. Alharbi and M.~Aldossary, ``Energy-efficient edge-fog-cloud architecture
  for iot-based smart agriculture environment,'' \emph{IEEE Access}, vol.~9,
  pp. 110\,480--110\,492, 2021.

\bibitem{Das2017CVPR}
A.~Das, M.~Degeling, X.~Wang, J.~Wang, N.~Sadeh, and M.~Satyanarayanan,
  ``Assisting users in a world full of cameras: A privacy-aware infrastructure
  for computer vision applications,'' in \emph{Proc. of IEEE Conference on
  Computer Vision and Pattern Recognition Workshops (CVPRW)}, 2017, pp.
  1387--1396.

\bibitem{Cheng2018CoRR}
C.-H. Cheng and I.~E. Olatunji, ``Harnessing constrained resources in service
  industry via video analytics,'' \emph{arXiv preprint arXiv:1807.00139}, 2018.

\bibitem{Xu2019ISBN}
M.~Xu, X.~Zhang, Y.~Liu, X.~Liu, and F.~X. Lin, ``Approximate query processing
  on autonomous cameras,'' 2019.

\bibitem{Wang2019IOTJ}
D.~{Wang}, P.~{Hu}, J.~{Du}, P.~{Zhou}, T.~{Deng}, and M.~{Hu}, ``Routing and
  scheduling for hybrid truck-drone collaborative parcel delivery with
  independent and truck-carried drones,'' \emph{IEEE Internet of Things
  Journal}, vol.~6, no.~6, pp. 10\,483--10\,495, 2019.

\bibitem{Funabashi2019RTSS}
Y.~{Funabashi}, I.~{Taniguchi}, and H.~{Tomiyama}, ``Work-in-progress: Routing
  of delivery drones with load-dependent flight speed,'' in \emph{2019 IEEE
  Real-Time Systems Symposium (RTSS)}, 2019, pp. 520--523.

\bibitem{Huang2020TITS}
H.~{Huang}, A.~V. {Savkin}, and C.~{Huang}, ``Reliable path planning for drone
  delivery using a stochastic time-dependent public transportation network,''
  \emph{IEEE Transactions on Intelligent Transportation Systems}, pp. 1--10,
  2020.

\bibitem{Long2018FIE}
R.~Long, T.~Tuna, and J.~Subhlok, ``Lecture video analytics as an instructional
  resource,'' in \emph{Proc. of IEEE Frontiers in Education Conference (FIE)},
  2018.

\bibitem{Jang2018SEC}
S.~Y. Jang, Y.~Lee, B.~Shin, and D.~Lee, ``Application-aware iot camera
  virtualization for video analytics edge computing,'' in \emph{Proc. of
  IEEE/ACM Symposium on Edge Computing (SEC)}, 2018, pp. 132--144.

\bibitem{Tarasov2018AIST}
A.~V. Tarasov and A.~V. Savchenko, ``Emotion recognition of a group of people
  in video analytics using deep off-the-shelf image embeddings,'' in
  \emph{Proc. of International Conference on Analysis of Images, Social
  Networks and Texts}.\hskip 1em plus 0.5em minus 0.4em\relax Springer, 2018,
  pp. 191--198.

\bibitem{Hu2019TPDS}
M.~{Hu}, L.~{Zhuang}, D.~{Wu}, Y.~{Zhou}, X.~{Chen}, and L.~{Xiao}, ``Learning
  driven computation offloading for asymmetrically informed edge computing,''
  \emph{IEEE Transactions on Parallel and Distributed Systems}, vol.~30, no.~8,
  pp. 1802--1815, Aug. 2019.

\bibitem{Wu2020PAAP_VBSSR}
R.~Wu, G.~Zhou, M.~Hu, and D.~Wu, ``Vbssr: Variable bitrate encoded video
  streaming with super-resolution on hpc education platform,'' in
  \emph{Parallel Architectures, Algorithms and Programming}, L.~Ning, V.~Chau,
  and F.~Lau, Eds.\hskip 1em plus 0.5em minus 0.4em\relax Singapore: Springer
  Singapore, 2021, pp. 224--235.

\bibitem{Wu2020PAAP}
R.~Wu, C.~Luo, M.~Hu, and D.~Wu, ``Inferring prerequisite relationships among
  learning resources for hpc education,'' in \emph{Parallel Architectures,
  Algorithms and Programming}, L.~Ning, V.~Chau, and F.~Lau, Eds.\hskip 1em
  plus 0.5em minus 0.4em\relax Singapore: Springer Singapore, 2021, pp.
  270--281.

\bibitem{Zhang2015MobiCom}
T.~Zhang, A.~Chowdhery, P.~V. Bahl, K.~Jamieson, and S.~Banerjee, ``The design
  and implementation of a wireless video surveillance system,'' in
  \emph{Proceedings of the 21st Annual International Conference on Mobile
  Computing and Networking (MobiCom)}.\hskip 1em plus 0.5em minus 0.4em\relax
  ACM, 2015, pp. 426--438.

\bibitem{Schindler2019MMM}
A.~Schindler, M.~Boyer, A.~Lindley, D.~Schreiber, and T.~Philipp, ``Large scale
  audio-visual video analytics platform for forensic investigations of
  terroristic attacks,'' in \emph{International Conference on Multimedia
  Modeling}.\hskip 1em plus 0.5em minus 0.4em\relax Springer, 2019, pp.
  106--119.

\bibitem{Khochare2019CCGRID}
A.~Khochare, S.~K. R., S.~R., and Y.~Simmhan, ``Dynamic scaling of video
  analytics for wide-area tracking in urban spaces,'' in \emph{IEEE/ACM
  International Symposium on Cluster, Cloud and Grid Computing(CCGrid)}, 2019.

\bibitem{Jain2019HotMobile}
S.~Jain, G.~Ananthanarayanan, J.~Jiang, Y.~Shu, and J.~E. Gonzalez, ``Scaling
  video analytics systems to large camera deployments,'' in \emph{Proceedings
  of the 20th International Workshop on Mobile Computing Systems and
  Applications (HotMobile)}, Feb. 2019.

\bibitem{Chowdhery2018SECON}
A.~Chowdhery and M.~Chiang, ``Model predictive compression for drone video
  analytics,'' in \emph{Proc. of IEEE International Conference on Sensing,
  Communication and Networking (SECON Workshops)}, 2018, pp. 1--5.

\bibitem{Wang2018SEC}
J.~Wang, Z.~Feng, Z.~Chen, S.~George, M.~Bala, P.~Pillai, S.-W. Yang, and
  M.~Satyanarayanan, ``Bandwidth-efficient live video analytics for drones via
  edge computing,'' in \emph{Proc. of IEEE/ACM Symposium on Edge Computing
  (SEC)}, 2018, pp. 159--173.

\bibitem{George2019HotMobile}
S.~George, J.~Wang, M.~Bala, T.~Eiszler, P.~Pillai, and M.~Satyanarayanan,
  ``Towards drone-sourced live video analytics for the construction industry,''
  in \emph{Proceedings of the 20th International Workshop on Mobile Computing
  Systems and Applications (HotMobile)}, 2019, pp. 3--8.

\bibitem{Lu2016SoCC_Optasia}
Y.~Lu, A.~Chowdhery, and S.~Kandula, ``Optasia: A relational platform for
  efficient large-scale video analytics,'' in \emph{Proc. of the Seventh ACM
  Symposium on Cloud Computing (SoCC)}, 2016, pp. 57--70.

\bibitem{Qiu2018IoTDI}
H.~Qiu, X.~Liu, S.~Rallapalli, A.~J. Bency, K.~Chan, R.~Urgaonkar,
  B.~Manjunath, and R.~Govindan, ``Kestrel: Video analytics for augmented
  multi-camera vehicle tracking,'' in \emph{Proc. of IEEE/ACM Third
  International Conference on Internet-of-Things Design and Implementation
  (IoTDI)}, 2018, pp. 48--59.

\bibitem{Chen2016SEC}
N.~Chen, Y.~Chen, S.~Song, C.-T. Huang, and X.~Ye, ``Smart urban surveillance
  using fog computing,'' in \emph{Proc. of IEEE/ACM Symposium on Edge Computing
  (SEC)}, 2016, pp. 95--96.

\bibitem{Chen2016BigMM}
N.~Chen, Y.~Chen, Y.~You, H.~Ling, P.~Liang, and R.~Zimmermann, ``Dynamic urban
  surveillance video stream processing using fog computing,'' in \emph{Proc. of
  IEEE second international conference on multimedia big data (BigMM)}, 2016,
  pp. 105--112.

\bibitem{Hu2021JNCA}
M.~Hu, X.~Luo, J.~Chen, Y.~C. Lee, Y.~Zhou, and D.~Wu, ``Virtual reality: A
  survey of enabling technologies and its applications in iot,'' \emph{Journal
  of Network and Computer Applications}, vol. 178, p. 102970, 2021.

\bibitem{Jain2015MobiSys_OverLay}
P.~Jain, J.~Manweiler, and R.~Roy~Choudhury, ``Overlay: Practical mobile
  augmented reality,'' in \emph{Proceedings of the 13th Annual International
  Conference on Mobile Systems, Applications, and Services (MobiSys)}.\hskip
  1em plus 0.5em minus 0.4em\relax ACM, 2015, pp. 331--344.

\bibitem{Jain2016CoNEXT}
------, ``Low bandwidth offload for mobile ar,'' in \emph{Proceedings of the
  12th International on Conference on emerging Networking EXperiments and
  Technologies (CoNEXT)}.\hskip 1em plus 0.5em minus 0.4em\relax ACM, 2016, pp.
  237--251.

\bibitem{Qiu2018MobiSys_AVR}
H.~Qiu, F.~Ahmad, F.~Bai, M.~Gruteser, and R.~Govindan, ``Avr: Augmented
  vehicular reality,'' in \emph{Proceedings of the 16th Annual International
  Conference on Mobile Systems, Applications, and Services (MobiSys)}.\hskip
  1em plus 0.5em minus 0.4em\relax ACM, 2018, pp. 81--95.

\bibitem{Liu2019MobiCom}
L.~Liu, H.~Li, and M.~Gruteser, ``Edge assisted real-time object detection for
  mobile augmented reality,'' in \emph{Proc. of ACM MobiCom}, 2019.

\bibitem{Zhang2017VR_ARNetwork}
W.~Zhang, B.~Han, and P.~Hui, ``On the networking challenges of mobile
  augmented reality,'' in \emph{Proceedings of the Workshop on Virtual Reality
  and Augmented Reality Network}.\hskip 1em plus 0.5em minus 0.4em\relax ACM,
  2017, pp. 24--29.

\bibitem{Ran2019HotNets}
X.~Ran, C.~Slocum, M.~Gorlatova, and J.~Chen, ``Sharear:
  Communication-efficient multi-user mobile augmented reality,'' in
  \emph{Proceedings of the 18th ACM Workshop on Hot Topics in Networks
  (HotNets)}.\hskip 1em plus 0.5em minus 0.4em\relax ACM, 2019, pp. 109--116.

\bibitem{Apicharttrisorn2019SenSys}
K.~Apicharttrisorn, X.~Ran, J.~Chen, S.~V. Krishnamurthy, and A.~K.
  Roy-Chowdhury, ``Frugal following: Power thrifty object detection and
  tracking for mobile augmented reality,'' in \emph{Proceedings of the 17th
  Conference on Embedded Networked Sensor Systems (SenSys)}, 2019, pp. 96--109.

\bibitem{Qiao2018InternetComputing_WebAR}
X.~Qiao, P.~Ren, S.~Dustdar, and J.~Chen, ``A new era for web ar with mobile
  edge computing,'' \emph{IEEE Internet Computing}, vol.~22, no.~4, pp. 46--55,
  2018.

\bibitem{Zhang2018HotMobile}
W.~Zhang, B.~Han, P.~Hui, V.~Gopalakrishnan, E.~Zavesky, and F.~Qian, ``{CARS}:
  collaborative augmented reality for socialization,'' in \emph{Proceedings of
  the 19th International Workshop on Mobile computing Systems \&
  Applications}.\hskip 1em plus 0.5em minus 0.4em\relax ACM, 2018, pp. 25--30.

\bibitem{Liu2016SEC_ParaDrop}
P.~{Liu}, D.~{Willis}, and S.~{Banerjee}, ``{ParaDrop}: Enabling lightweight
  multi-tenancy at the network’s extreme edge,'' in \emph{Proc. of IEEE/ACM
  Symposium on Edge Computing (SEC)}, Oct. 2016, pp. 1--13.

\bibitem{King2020EdgeSum}
J.~{King}, L.~{Huang}, D.~{Wu}, Y.~{Zhou}, and Y.~C. {Lee}, ``{EdgeSum}:
  Edge-based video summarization with dash cams,'' in \emph{Proc. of IEEE
  International Conference on Cloud Engineering (IC2E)}, 2020, pp. 40--48.

\bibitem{Wang2017SmartIOT}
J.~Wang, J.~Pan, and F.~Esposito, ``Elastic urban video surveillance system
  using edge computing,'' in \emph{Proceedings of the Workshop on Smart
  Internet of Things (SmartIOT)}.\hskip 1em plus 0.5em minus 0.4em\relax ACM,
  2017, p.~7.

\bibitem{Liu2018EdgeSys_EdgeEye}
P.~Liu, B.~Qi, and S.~Banerjee, ``Edgeeye: An edge service framework for
  real-time intelligent video analytics,'' in \emph{Proceedings of the 1st
  International Workshop on Edge Systems, Analytics and Networking
  (EdgeSys)}.\hskip 1em plus 0.5em minus 0.4em\relax ACM, 2018, pp. 1--6.

\bibitem{Stone2019SECON}
T.~Stone, N.~Stone, P.~Jain, Y.~Jiang, K.-H. Kim, and S.~Nelakuditi1, ``Towards
  scalable video analytics at the edge,'' in \emph{Proc. of IEEE International
  Conference on Sensing, Communication, and Networking (SECON)}, 2019.

\bibitem{Luo2018SEC_EdgeBox}
B.~Luo, S.~Tan, Z.~Yu, and W.~Shi, ``{EdgeBox}: Live edge video analytics for
  near real-time event detection,'' in \emph{Proc. of IEEE/ACM Symposium on
  Edge Computing (SEC)}, 2018, pp. 347--348.

\bibitem{Dao2017ICDCS}
T.~Dao, K.~Khalil, A.~K. Roy-Chowdhury, S.~V. Krishnamurthy, and L.~Kaplan,
  ``Energy efficient object detection in camera sensor networks,'' in
  \emph{Proc. of IEEE 37th International Conference on Distributed Computing
  Systems (ICDCS)}, 2017, pp. 1208--1218.

\bibitem{Ali2018ICFEC}
M.~Ali, A.~Anjum, M.~U. Yaseen, A.~R. Zamani, D.~Balouek-Thomert, O.~Rana, and
  M.~Parashar, ``Edge enhanced deep learning system for large-scale video
  stream analytics,'' in \emph{Proc. of IEEE 2nd International Conference on
  Fog and Edge Computing (ICFEC)}, 2018, pp. 1--10.

\bibitem{Ananthanarayanan2017computer}
G.~Ananthanarayanan, P.~Bahl, P.~Bod{\'\i}k, K.~Chintalapudi, M.~Philipose,
  L.~Ravindranath, and S.~Sinha, ``Real-time video analytics: The killer app
  for edge computing,'' \emph{IEEE Computer}, vol.~50, no.~10, pp. 58--67,
  2017.

\bibitem{Perala2018ISCAS}
S.~S.~N. Perala, I.~Galanis, and I.~Anagnostopoulos, ``Fog computing and
  efficient resource management in the era of internet-of-video things
  ({IoVT}),'' in \emph{Proc. of IEEE International Symposium on Circuits and
  Systems (ISCAS)}, 2018, pp. 1--5.

\bibitem{Drolia2017ICDCS_Cachier}
U.~Drolia, K.~Guo, J.~Tan, R.~Gandhi, and P.~Narasimhan, ``Cachier:
  Edge-caching for recognition applications,'' in \emph{Proc. of IEEE 37th
  International Conference on Distributed Computing Systems (ICDCS)}, 2017, pp.
  276--286.

\bibitem{Wang2019HotCloud}
Y.~Wang, W.~Wang, J.~Zhang, J.~Jiang, and K.~Chen, ``Bridging the edge-cloud
  barrier for real-time advanced vision analytics,'' in \emph{Proc. of 11th
  USENIX Workshop on Hot Topics in Cloud Computing (HotCloud)}, Renton, WA,
  July 2019.

\bibitem{Canel2019SysML}
C.~Canel, T.~Kim, G.~Zhou, C.~Li, H.~Lim, D.~G. Andersen, M.~Kaminsky, and
  S.~R. Dulloor, ``Scaling video analytics on constrained edge nodes,''
  \emph{arXiv preprint arXiv:1905.13536}, 2019.

\bibitem{Ran2018INFOCOM_DeepDecision}
X.~Ran, H.~Chen, X.~Zhu, Z.~Liu, and J.~Chen, ``Deepdecision: A mobile deep
  learning framework for edge video analytics,'' in \emph{Proc. of IEEE
  Conference on Computer Communications (INFOCOM)}, 2018, pp. 1421--1429.

\bibitem{Guo2019TMM}
Y.~Guo, B.~Zou, J.~Ren, Q.~Liu, D.~Zhang, and Y.~Zhang, ``Distributed and
  efficient object detection via interactions among devices, edge and cloud,''
  \emph{IEEE Transactions on Multimedia}, 2019.

\bibitem{Zhang2016SEC}
Q.~Zhang, Z.~Yu, W.~Shi, and H.~Zhong, ``Demo abstract: Evaps: Edge video
  analysis for public safety,'' in \emph{Proc. of IEEE/ACM Symposium on Edge
  Computing (SEC)}, 2016, pp. 121--122.

\bibitem{Long2017TMM}
C.~Long, Y.~Cao, T.~Jiang, and Q.~Zhang, ``Edge computing framework for
  cooperative video processing in multimedia iot systems,'' \emph{IEEE
  Transactions on Multimedia}, vol.~20, no.~5, pp. 1126--1139, 2017.

\bibitem{Zhang2018ICPP}
C.~Zhang, Q.~Cao, H.~Jiang, W.~Zhang, J.~Li, and J.~Yao, ``Ffs-va: A fast
  filtering system for large-scale video analytics,'' in \emph{Proceedings of
  the 47th International Conference on Parallel Processing (ICPP)}.\hskip 1em
  plus 0.5em minus 0.4em\relax ACM, 2018, p.~85.

\bibitem{Dao2017MASS}
T.~Dao, A.~Roy-Chowdhury, N.~Nasrabadi, S.~V. Krishnamurthy, P.~Mohapatra, and
  L.~M. Kaplan, ``Accurate and timely situation awareness retrieval from a
  bandwidth constrained camera network,'' in \emph{Proc. of IEEE 14th
  International Conference on Mobile Ad Hoc and Sensor Systems (MASS)}, 2017,
  pp. 416--425.

\bibitem{Wang2016TMM}
S.~Wang, X.~Zhang, X.~Liu, J.~Zhang, S.~Ma, and W.~Gao, ``Utility-driven
  adaptive preprocessing for screen content video compression,'' \emph{IEEE
  Transactions on Multimedia}, vol.~19, no.~3, pp. 660--667, 2016.

\bibitem{Chen2020NOSSDAV}
J.~Chen, M.~Hu, Z.~Luo, Z.~Wang, and D.~Wu, ``{SR360}: boosting 360-degree
  video streaming with super-resolution,'' in \emph{Proceedings of the 30th ACM
  Workshop on Network and Operating Systems Support for Digital Audio and Video
  (NOSSDAV)}, 2020, pp. 1--6.

\bibitem{Jiang2018ATC}
A.~H. Jiang, D.~L.-K. Wong, C.~Canel, L.~Tang, I.~Misra, M.~Kaminsky, M.~A.
  Kozuch, P.~Pillai, D.~G. Andersen, and G.~R. Ganger, ``Mainstream: Dynamic
  stem-sharing for multi-tenant video processing,'' in \emph{Proc. of USENIX
  Annual Technical Conference (ATC)}, 2018, pp. 29--42.

\bibitem{Lu18INFOCOM}
Z.~Lu, K.~S. Chan, and T.~La~Porta, ``A computing platform for video
  crowdprocessing using deep learning,'' in \emph{Proc. of IEEE Conference on
  Computer Communications (INFOCOM)}, 2018, pp. 1430--1438.

\bibitem{Lu18TMC_CrowdVision}
Z.~Lu, K.~Chan, S.~Pu, and T.~La~Porta, ``Crowdvision: A computing platform for
  video crowdprocessing using deep learning,'' \emph{IEEE Transactions on
  Mobile Computing}, vol.~18, no.~7, pp. 1513--1526, 2018.

\bibitem{tyolo}
\BIBentryALTinterwordspacing
``Ameema zainab. real-time object detection,'' 2017. [Online]. Available:
  \url{https://blog.mindorks.com/detection-on-android-using-tensorflow-a3f6fe423349}
\BIBentrySTDinterwordspacing

\bibitem{pulli2012real}
K.~Pulli, A.~Baksheev, K.~Kornyakov, and V.~Eruhimov, ``Real-time computer
  vision with opencv,'' \emph{Communications of the ACM}, vol.~55, no.~6, pp.
  61--69, 2012.

\bibitem{howard2017mobilenets}
A.~G. Howard, M.~Zhu, B.~Chen, D.~Kalenichenko, W.~Wang, T.~Weyand,
  M.~Andreetto, and H.~Adam, ``{MobileNets}: Efficient convolutional neural
  networks for mobile vision applications,'' \emph{arXiv preprint
  arXiv:1704.04861}, 2017.

\bibitem{Rippel2019ICCV}
O.~Rippel, S.~Nair, C.~Lew, S.~Branson, A.~G. Anderson, and L.~Bourdev,
  ``Learned video compression,'' in \emph{Proceedings of the IEEE International
  Conference on Computer Vision (ICCV)}, 2019, pp. 3454--3463.

\bibitem{Fouladi2017nsdi}
S.~Fouladi, R.~S. Wahby, B.~Shacklett, K.~V. Balasubramaniam, W.~Zeng,
  R.~Bhalerao, A.~Sivaraman, G.~Porter, and K.~Winstein, ``Encoding, fast and
  slow: Low-latency video processing using thousands of tiny threads,'' in
  \emph{Proc. of 14th USENIX Symposium on Networked Systems Design and
  Implementation (NSDI)}, 2017, pp. 363--376.

\bibitem{Yang2018Access}
X.~Yang, W.~Wu, K.~Liu, P.~W. Kim, A.~K. Sangaiah, and G.~Jeon,
  ``Multi-semi-couple super-resolution method for edge computing,'' \emph{IEEE
  Access}, vol.~6, pp. 5511--5520, 2018.

\bibitem{Guo2019sensors}
J.~Guo, X.~Gong, W.~Wang, X.~Que, and J.~Liu, ``{SASRT}: semantic-aware
  super-resolution transmission for adaptive video streaming over wireless
  multimedia sensor networks,'' \emph{Sensors}, vol.~19, no.~14, p. 3121, 2019.

\bibitem{Uddin2019Symmetry}
M.~A. Uddin, A.~Alam, N.~A. Tu, M.~S. Islam, and Y.-K. Lee, ``{SIAT}: A
  distributed video analytics framework for intelligent video surveillance,''
  \emph{Symmetry}, vol.~11, no.~7, p. 911, 2019.

\bibitem{lowe2004distinctive}
D.~G. Lowe, ``Distinctive image features from scale-invariant keypoints,''
  \emph{International journal of computer vision}, vol.~60, no.~2, pp. 91--110,
  2004.

\bibitem{Kang2017PVLDB_NoScope}
D.~Kang, J.~Emmons, F.~Abuzaid, P.~Bailis, and M.~Zaharia, ``Noscope:
  optimizing neural network queries over video at scale,'' \emph{Proceedings of
  the VLDB Endowment}, vol.~10, no.~11, pp. 1586--1597, 2017.

\bibitem{Lu17MM}
Z.~Lu, S.~Rallapalli, K.~Chan, and T.~La~Porta, ``Modeling the resource
  requirements of convolutional neural networks on mobile devices,'' in
  \emph{Proceedings of the 25th ACM international conference on Multimedia
  (MM)}, 2017, pp. 1663--1671.

\bibitem{Lu2020ToN_NetVision}
Z.~{Lu}, K.~{Chan}, R.~{Urgaonkar}, S.~{Pu}, and T.~{La Porta}, ``Netvision:
  On-demand video processing in wireless networks,'' \emph{IEEE/ACM
  Transactions on Networking}, vol.~28, no.~1, pp. 196--209, 2020.

\bibitem{Huang18MM_QARC}
T.~Huang, R.-X. Zhang, C.~Zhou, and L.~Sun, ``Qarc: Video quality aware rate
  control for real-time video streaming based on deep reinforcement learning,''
  in \emph{Proc. of ACM Conference on Multimedia (MM)}, 2018, pp. 1208--1216.

\bibitem{he2016identity}
K.~He, X.~Zhang, S.~Ren, and J.~Sun, ``Identity mappings in deep residual
  networks,'' in \emph{Proc. of European Conference on Computer Vision
  (ECCV)}.\hskip 1em plus 0.5em minus 0.4em\relax Springer, 2016, pp. 630--645.

\bibitem{simonyan2014very}
K.~Simonyan and A.~Zisserman, ``Very deep convolutional networks for
  large-scale image recognition,'' \emph{arXiv preprint arXiv:1409.1556}, 2014.

\bibitem{zoph2018learning}
B.~Zoph, V.~Vasudevan, J.~Shlens, and Q.~V. Le, ``Learning transferable
  architectures for scalable image recognition,'' in \emph{Proceedings of the
  IEEE conference on computer vision and pattern recognition (CVPR)}, 2018, pp.
  8697--8710.

\bibitem{alexnet}
A.~Krizhevsky, I.~Sutskever, and G.~E. Hinton, ``Imagenet classification with
  deep convolutional neural networks,'' in \emph{Proc. of Advances in Neural
  Information Processing Systems (NIPS)}, 2012, pp. 1097--1105.

\bibitem{gglnet}
C.~Szegedy, W.~Liu, Y.~Jia, P.~Sermanet, S.~Reed, D.~Anguelov, D.~Erhan,
  V.~Vanhoucke, and A.~Rabinovich, ``Going deeper with convolutions,'' in
  \emph{Proceedings of the IEEE conference on computer vision and pattern
  recognition (CVPR)}, 2015, pp. 1--9.

\bibitem{jetsondevkit}
\emph{{NVIDIA} Corporation,
  https://developer.nvidia.com/embedded/buy/jetson-devkit}, 2020.

\bibitem{Yaseen2019TSMC}
M.~U. Yaseen, A.~Anjum, O.~Rana, and N.~Antonopoulos, ``Deep learning
  hyper-parameter optimization for video analytics in clouds,'' \emph{IEEE
  Transactions on Systems, Man, and Cybernetics: Systems}, vol.~49, no.~1, pp.
  253--264, 2018.

\bibitem{Han2016MobiSys}
S.~Han, H.~Shen, M.~Philipose, S.~Agarwal, A.~Wolman, and A.~Krishnamurthy,
  ``Mcdnn: An approximation-based execution framework for deep stream
  processing under resource constraints,'' in \emph{Proceedings of the 14th
  Annual International Conference on Mobile Systems, Applications, and
  Services}, 2016, pp. 123--136.

\bibitem{Felemban2020TMC}
N.~Felemban, F.~Mehmeti, H.~Khamfroush, Z.~Lu, S.~Rallapalli, K.~S. Chan, and
  T.~La~Porta, ``Picsys: Energy-efficient fast image search on distributed
  mobile networks,'' \emph{IEEE Transactions on Mobile Computing}, 2019.

\bibitem{Kang2017ASPLOS}
Y.~Kang, J.~Hauswald, C.~Gao, A.~Rovinski, T.~Mudge, J.~Mars, and L.~Tang,
  ``Neurosurgeon: Collaborative intelligence between the cloud and mobile
  edge,'' in \emph{ACM SIGARCH Computer Architecture News}, vol.~45,
  no.~1.\hskip 1em plus 0.5em minus 0.4em\relax ACM, 2017, pp. 615--629.

\bibitem{Emmons2019HotEdgeVideo}
J.~Emmons, S.~Fouladi, G.~Ananthanarayanan, S.~Venkataraman, S.~Savarese, and
  K.~Winstein, ``Cracking open the {DNN} black-box: Video analytics with {DNNs}
  across the camera-cloud boundary,'' in \emph{Proceedings of the Workshop on
  Hot Topics in Video Analytics and Intelligent Edges (HotEdgeVideo)}, 2019,
  pp. 27--32.

\bibitem{Jiang2018SIGCOMM_Chameleon}
J.~Jiang, G.~Ananthanarayanan, P.~Bodik, S.~Sen, and I.~Stoica, ``Chameleon:
  scalable adaptation of video analytics,'' in \emph{Proceedings of the ACM
  Special Interest Group on Data Communication (SIGCOMM)}, 2018, pp. 253--266.

\bibitem{O'Gorman2018ICPR}
L.~O'Gorman and X.~Wang, ``Balancing video analytics processing and bandwidth
  for edge-cloud networks,'' in \emph{Proc. of 24th International Conference on
  Pattern Recognition (ICPR)}, 2018, pp. 2618--2623.

\bibitem{Wang2018IC-NIDC}
J.~Wang, Y.~Hu, H.~Li, and G.~Shou, ``A lightweight edge computing platform
  integration video services,'' in \emph{Proc. of International Conference on
  Network Infrastructure and Digital Content (IC-NIDC)}.\hskip 1em plus 0.5em
  minus 0.4em\relax IEEE, 2018, pp. 183--187.

\bibitem{xu2018ATC}
R.~Xu, J.~Koo, R.~Kumar, P.~Bai, S.~Mitra, S.~Misailovic, and S.~Bagchi,
  ``Videochef: efficient approximation for streaming video processing
  pipelines,'' in \emph{Proc. of USENIX Annual Technical Conference (ATC)},
  2018, pp. 43--56.

\bibitem{Fu2019ATC_EdgeWise}
X.~Fu, T.~Ghaffar, J.~C. Davis, and D.~Lee, ``Edgewise: a better stream
  processing engine for the edge,'' in \emph{Proc. of USENIX Annual Technical
  Conference (ATC)}, 2019, pp. 929--946.

\bibitem{Shen2017CVPR}
H.~Shen, S.~Han, M.~Philipose, and A.~Krishnamurthy, ``Fast video
  classification via adaptive cascading of deep models,'' in \emph{Proceedings
  of the IEEE conference on computer vision and pattern recognition (CVPR)},
  2017, pp. 3646--3654.

\bibitem{Valls2018MILCOM}
V.~Valls, H.~Kwon, T.~LaPorta, S.~Stein, and L.~Tassiulas, ``On the design of
  resource allocation algorithms for low-latency video analytics,'' in
  \emph{Proc. of IEEE Military Communications Conference (MILCOM)}, 2018, pp.
  1--6.

\bibitem{Rao2017ICCV}
Y.~Rao, J.~Lu, and J.~Zhou, ``Attention-aware deep reinforcement learning for
  video face recognition,'' in \emph{Proceedings of the IEEE International
  Conference on Computer Vision (ICCV)}, 2017, pp. 3931--3940.

\bibitem{Supancic2017ICCV}
J.~Supancic~III and D.~Ramanan, ``Tracking as online decision-making: Learning
  a policy from streaming videos with reinforcement learning,'' in
  \emph{Proceedings of the IEEE International Conference on Computer Vision
  (ICCV)}, 2017, pp. 322--331.

\bibitem{Zhan2016MSN}
Y.~Zhan, Y.~Liu, Y.~Xia, L.~Yan, F.~Li, N.~Zhou, and H.~Wu, ``Incentive
  mechanism for crowdsourced mobile video offloading,'' in \emph{Proc. of 12th
  International Conference on Mobile Ad-Hoc and Sensor Networks (MSN)}.\hskip
  1em plus 0.5em minus 0.4em\relax IEEE, 2016, pp. 279--283.

\bibitem{Wu2017TMM}
D.~Wu, J.~Yan, H.~Wang, D.~Wu, and R.~Wang, ``Social attribute aware incentive
  mechanism for device-to-device video distribution,'' \emph{IEEE Transactions
  on Multimedia}, vol.~19, no.~8, pp. 1908--1920, 2017.

\bibitem{Hu2020TPDS}
M.~{Hu}, Z.~{Xie}, D.~{Wu}, Y.~{Zhou}, X.~{Chen}, and L.~{Xiao},
  ``Heterogeneous edge offloading with incomplete information: A minority game
  approach,'' \emph{IEEE Transactions on Parallel and Distributed Systems},
  vol.~31, no.~9, pp. 2139--2154, 2020.

\bibitem{Poostchi2017CoRR}
M.~Poostchi, K.~Palaniappan, D.~Li, M.~Becchi, F.~Bunyak, and G.~Seetharaman,
  ``Fast integral histogram computations on gpu for real-time video
  analytics,'' \emph{arXiv preprint arXiv:1711.01919}, 2017.

\bibitem{Xu2019arXiv_2}
M.~Xu, T.~Xu, Y.~Liu, X.~Liu, G.~Huang, and F.~X. Lin, ``Supporting video
  queries on zero-streaming cameras,'' \emph{arXiv preprint arXiv:1904.12342},
  2019.

\bibitem{Huang2017sosp_SVE}
Q.~Huang, P.~Ang, P.~Knowles, T.~Nykiel, I.~Tverdokhlib, A.~Yajurvedi,
  P.~Dapolito~IV, X.~Yan, M.~Bykov, C.~Liang \emph{et~al.}, ``Sve: Distributed
  video processing at facebook scale,'' in \emph{Proceedings of the 26th
  Symposium on Operating Systems Principles (SOSP)}, 2017, pp. 87--103.

\bibitem{jetsonnvidiatx2}
\emph{{NVIDIA} Corporation, https://developer.nvidia.com/embedded/jetson-tx2},
  2020.

\bibitem{jetsonnvidianano}
\emph{{NVIDIA} Corporation,
  https://www.nvidia.com/en-au/autonomous-machines/jetson-store/}, 2020.

\bibitem{awssnowball}
\emph{{Amazon} Corporation, https://aws.amazon.com/snowball/}, 2020.

\bibitem{azureiot}
\emph{Azure Corporation,
  https://developer.qualcomm.com/hardware/vision-ai-development-kit}, 2020.

\bibitem{intelnuc}
\emph{Intel Corporation, https://simplynuc.co.uk/aidevkit}, 2020.

\bibitem{raspi}
\emph{Raspberry, https://www.raspberrypi.org/}, 2020.

\bibitem{onionomega}
\emph{Onion, https://onion.io/}, 2020.

\bibitem{orangepi}
\emph{Orange Pi, http://www.orangepi.org/}, 2020.

\bibitem{bananapi}
\emph{Orange Pi, http://www.http://www.banana-pi.org/}, 2020.

\bibitem{rock}
\emph{Rock, https://www.pine64.org/devices/single-board-computers/rock64/},
  2020.

\bibitem{arduino}
\emph{Arduino, https://www.arduino.cc/}, 2020.

\bibitem{tinkerboard}
\emph{Tinker, https://www.asus.com/au/Single-Board-Computer/Tinker-Board/},
  2020.

\bibitem{odroid}
\emph{Odroid, https://www.hardkernel.com/}, 2020.

\bibitem{pin64}
\emph{Pine64,
  https://www.pine64.org/devices/single-board-computers/pine-a64-lts/}, 2020.

\bibitem{cubieboard}
\emph{Cubieboard Board, http://cubieboard.org/}, 2020.

\bibitem{beagleboard}
\emph{Beagle Board, https://beagleboard.org/}, 2020.

\bibitem{lattepanda}
\emph{LattePanda, https://www.dfrobot.com/}, 2020.

\bibitem{udoobolt}
\emph{Udoo Bolt, https://www.udoo.org/udoo-bolt/}, 2020.

\bibitem{librecomputer}
\emph{Liber Computer, https://libre.computer/products/boards/}, 2020.

\bibitem{nanopi}
\emph{Nanopi, https://www.friendlyarm.com/}, 2020.

\bibitem{opencv}
\emph{OpenCV, https://opencv.org/}, 2020.

\bibitem{simplecv}
\emph{SimpleCV, http://simplecv.org/}, 2020.

\bibitem{scikitimage}
\emph{Scikit-image, https://scikit-image.org/}, 2020.

\bibitem{accordnetframework}
\emph{accordnetframe, http://accord-framework.net/}, 2020.

\bibitem{boofcv}
\emph{BoofCV, http://boofcv.org/}, 2020.

\bibitem{fastcv}
\emph{FastCV, https://developer.qualcomm.com/forums/software/fastcv}, 2020.

\bibitem{matlab}
\emph{MSTLAB, https://www.mathworks.com/products/matlab.html}, 2020.

\bibitem{deepface}
\emph{DeepFace, https://github.com/serengil/deepface}, 2020.

\bibitem{schroff2015facenet}
F.~Schroff, D.~Kalenichenko, and J.~Philbin, ``Facenet: A unified embedding for
  face recognition and clustering,'' in \emph{Proceedings of the IEEE
  conference on computer vision and pattern recognition}, 2015, pp. 815--823.

\bibitem{vggface}
O.~M. Parkhi, A.~Vedaldi, and A.~Zisserman, ``Deep face recognition,'' in
  \emph{British Machine Vision Conference}, 2015.

\bibitem{pcl}
\emph{Point Cloud Library (PCL), https://pointclouds.org/}, 2020.

\bibitem{nvidialib1}
\emph{NVIDIA CUDA-X , https://developer.nvidia.com/gpu-accelerated-libraries},
  2020.

\bibitem{nvidialib2}
\emph{NVIDIA Performance Primitives, https://developer.nvidia.com/npp}, 2020.

\bibitem{detectron2}
Y.~Wu, A.~Kirillov, F.~Massa, W.-Y. Lo, and R.~Girshick, ``Detectron2,''
  \url{https://github.com/facebookresearch/detectron2}, 2019.

\bibitem{yolov3}
J.~Redmon and A.~Farhadi, ``Yolov3: An incremental improvement,'' \emph{arXiv
  preprint arXiv:1804.02767}, 2018.

\bibitem{liu2018progressive}
C.~Liu, B.~Zoph, M.~Neumann, J.~Shlens, W.~Hua, L.-J. Li, L.~Fei-Fei,
  A.~Yuille, J.~Huang, and K.~Murphy, ``Progressive neural architecture
  search,'' in \emph{Proceedings of the European Conference on Computer Vision
  (ECCV)}, 2018, pp. 19--34.

\bibitem{kerasapi}
\emph{Keras, https://keras.io/}, 2020.

\bibitem{cafeapi}
\emph{Caffe, https://caffe.berkeleyvision.org/}, 2020.

\bibitem{pytorch}
\emph{PyTorch, https://pytorch.org/}, 2020.

\bibitem{albumentations}
\emph{Albumentations, https://albumentations.ai/}, 2020.

\bibitem{openvino}
\emph{OpenVINO, https://docs.openvino.ai/latest/index.html}, 2020.

\bibitem{tensorflow}
\emph{TensorFlow, https://www.tensorflow.org/}, 2020.

\bibitem{Kim2014ICUAS}
Y.~Kim, J.~Jo, and S.~Shrestha, ``A server-based real-time privacy protection
  scheme against video surveillance by unmanned aerial systems,'' in
  \emph{Proc. of International Conference on Unmanned Aircraft Systems
  (ICUAS)}.\hskip 1em plus 0.5em minus 0.4em\relax IEEE, 2014, pp. 684--691.

\bibitem{Wang2017MMSys}
J.~Wang, B.~Amos, A.~Das, P.~Pillai, N.~Sadeh, and M.~Satyanarayanan, ``A
  scalable and privacy-aware iot service for live video analytics,'' in
  \emph{Proceedings of the 8th ACM on Multimedia Systems Conference (MMSys)},
  2017, pp. 38--49.

\bibitem{Wang2018TOMCCAP}
------, ``Enabling live video analytics with a scalable and privacy-aware
  framework,'' \emph{ACM Transactions on Multimedia Computing, Communications,
  and Applications}, vol.~14, no.~3, p.~64, 2018.

\bibitem{Wang2019Access}
X.~Wang, H.~Xue, X.~Liu, and Q.~Pei, ``A privacy-preserving edge
  computation-based face verification system for user authentication,''
  \emph{IEEE Access}, vol.~7, pp. 14\,186--14\,197, 2019.

\bibitem{Li2018IOTJ}
X.~Li, S.~Liu, F.~Wu, S.~Kumari, and J.~J. Rodrigues, ``Privacy preserving data
  aggregation scheme for mobile edge computing assisted iot applications,''
  \emph{IEEE Internet of Things Journal}, vol.~6, no.~3, pp. 4755--4763, 2018.

\bibitem{Zarepour16PerCom}
E.~Zarepour, M.~Hosseini, S.~S. Kanhere, and A.~Sowmya, ``A context-based
  privacy preserving framework for wearable visual lifeloggers,'' in
  \emph{Proc. of IEEE International Conference on Pervasive Computing and
  Communication Workshops (PerCom Workshops)}, 2016, pp. 1--4.

\bibitem{wang2019arXiv}
H.~Wang, S.~Xie, and Y.~Hong, ``{VideoDP}: A universal platform for video
  analytics with differential privacy,'' \emph{arXiv preprint
  arXiv:1909.08729}, 2019.

\bibitem{xu2018Communications_Magazine}
C.~Xu, J.~Ren, D.~Zhang, and Y.~Zhang, ``Distilling at the edge: A local
  differential privacy obfuscation framework for iot data analytics,''
  \emph{IEEE Communications Magazine}, vol.~56, no.~8, pp. 20--25, 2018.

\bibitem{mao2018USENIX}
Y.~Mao, S.~Yi, Q.~Li, J.~Feng, F.~Xu, and S.~Zhong, ``A privacy-preserving deep
  learning approach for face recognition with edge computing,'' in \emph{Proc.
  of USENIX Workshop Hot Topics Edge Computing (HotEdge)}, 2018, pp. 1--6.

\bibitem{Jiang17DSC}
L.~Jiang, C.~Xu, X.~Wang, B.~Luo, and H.~Wang, ``Secure outsourcing sift:
  Efficient and privacy-preserving image feature extraction in the encrypted
  domain,'' \emph{IEEE Transactions on Dependable and Secure Computing}, 2017.

\bibitem{Khazbak2019IoTDI}
Y.~Khazbak, J.~Qiu, T.~Tan, and G.~Cao, ``Targetfinder: privacy preserving
  target search through iot cameras,'' in \emph{Proceedings of the
  International Conference on Internet of Things Design and Implementation
  (IoTDI)}.\hskip 1em plus 0.5em minus 0.4em\relax ACM, 2019, pp. 213--224.

\bibitem{Li2019FCS}
X.~Li, J.~Li, S.~Yiu, C.~Gao, and J.~Xiong, ``Privacy-preserving edge-assisted
  image retrieval and classification in iot,'' \emph{Frontiers of Computer
  Science}, vol.~13, no.~5, pp. 1136--1147, 2019.

\bibitem{akkaya2019CNS}
a.~Akkaya, V.~Baboolal, N.~Saputro, S.~Uluagac, and H.~Menouar,
  ``Privacy-preserving control of video transmissions for drone-based
  intelligent transportation systems,'' in \emph{Proc. of IEEE Conference on
  Communications and Network Security (CNS)}, 2019, pp. 1--7.

\bibitem{sada2019DASC}
A.~B. Sada, M.~A. Bouras, J.~Ma, H.~Runhe, and H.~Ning, ``A distributed video
  analytics architecture based on edge-computing and federated learning,'' in
  \emph{Proc. of IEEE Intl Conf on Dependable, Autonomic and Secure Computing,
  Intl Conf on Pervasive Intelligence and Computing, Intl Conf on Cloud and Big
  Data Computing, Intl Conf on Cyber Science and Technology Congress
  (DASC/PiCom/CBDCom/CyberSciTech)}, 2019, pp. 215--220.

\bibitem{chen2019GLOBECOM}
M.~Chen, O.~Semiari, W.~Saad, X.~Liu, and C.~Yin, ``Federated deep learning for
  immersive virtual reality over wireless networks,'' in \emph{Proc. of IEEE
  Global Communications Conference (GLOBECOM)}, 2019, pp. 1--6.

\bibitem{liu2020arXiv}
Y.~Liu, A.~Huang, Y.~Luo, H.~Huang, Y.~Liu, Y.~Chen, L.~Feng, T.~Chen, H.~Yu,
  and Q.~Yang, ``Fedvision: An online visual object detection platform powered
  by federated learning,'' \emph{arXiv preprint arXiv:2001.06202}, 2020.

\bibitem{xu2019arXiv_3}
D.~Xu, M.~Zheng, L.~Jiang, C.~Gu, R.~Tan, and P.~Cheng, ``Lightweight and
  unobtrusive privacy preservation for remote inference via edge data
  obfuscation,'' \emph{arXiv preprint arXiv:1912.09859}, 2019.

\bibitem{osia2020IOT}
S.~A. {Osia}, A.~{Shahin Shamsabadi}, S.~{Sajadmanesh}, A.~{Taheri},
  K.~{Katevas}, H.~R. {Rabiee}, N.~D. {Lane}, and H.~{Haddadi}, ``A hybrid deep
  learning architecture for privacy-preserving mobile analytics,'' \emph{IEEE
  Internet of Things Journal}, vol.~7, no.~5, pp. 4505--4518, 2020.

\bibitem{Hu2021Springer}
M.~Hu, Y.~Fu, and D.~Wu, ``Privacy-preserving edge video analytics,'' in
  \emph{Fog/Edge Computing For Security, Privacy, and Applications}.\hskip 1em
  plus 0.5em minus 0.4em\relax Springer, 2021, pp. 171--190.

\bibitem{Nigel2016HotMobile}
N.~Davies, N.~Taft, M.~Satyanarayanan, S.~Clinch, and B.~Amos, ``Privacy
  mediators: Helping iot cross the chasm,'' in \emph{Proceedings of the 17th
  International Workshop on Mobile Computing Systems and Applications
  (HotMobile)}.\hskip 1em plus 0.5em minus 0.4em\relax ACM, 2016, pp. 39--44.

\bibitem{Wang2020JPDC}
T.~Wang, Y.~Mei, W.~Jia, X.~Zheng, G.~Wang, and M.~Xie, ``Edge-based
  differential privacy computing for sensor--cloud systems,'' \emph{Journal of
  Parallel and Distributed Computing}, vol. 136, pp. 75--85, 2020.

\bibitem{Xiao2020Sensors}
D.~Xiao, M.~Li, and H.~Zheng, ``Smart privacy protection for big video data
  storage based on hierarchical edge computing,'' \emph{Sensors}, vol.~20,
  no.~5, p. 1517, 2020.

\bibitem{melis2019sp}
L.~Melis, C.~Song, E.~De~Cristofaro, and V.~Shmatikov, ``Exploiting unintended
  feature leakage in collaborative learning,'' in \emph{Proc. of IEEE Symposium
  on Security and Privacy (SP)}, 2019, pp. 691--706.

\bibitem{FHE_Gentry2009}
C.~Gentry, ``Fully homomorphic encryption using ideal lattices,'' in
  \emph{Proceedings of the forty-first Annual ACM Symposium on Theory of
  Computing (STOC)}, 2009, pp. 169--178.

\bibitem{ma2018INFOCOM}
X.~Ma, B.~Zhu, T.~Zhang, S.~Cao, H.~Jin, and D.~Zou, ``Efficient
  privacy-preserving motion detection for hevc compressed video in cloud video
  surveillance,'' in \emph{Proc. of IEEE Conference on Computer Communications
  Workshops (INFOCOM WKSHPS)}, 2018, pp. 813--818.

\bibitem{chi2018arXiv}
J.~Chi, E.~Owusu, X.~Yin, T.~Yu, W.~Chan, P.~Tague, and Y.~Tian, ``Privacy
  partitioning: Protecting user data during the deep learning inference
  phase,'' \emph{arXiv preprint arXiv:1812.02863}, 2018.

\bibitem{osia2017privacy}
S.~A. Osia, A.~S. Shamsabadi, A.~Taheri, K.~Katevas, H.~R. Rabiee, N.~D. Lane,
  and H.~Haddadi, ``Privacy-preserving deep inference for rich user data on the
  cloud,'' \emph{arXiv preprint arXiv:1710.01727}, 2017.

\bibitem{chopra2005learning}
S.~Chopra, R.~Hadsell, and Y.~LeCun, ``Learning a similarity metric
  discriminatively, with application to face verification,'' in \emph{Proc. of
  IEEE Computer Society Conference on Computer Vision and Pattern Recognition
  (CVPR)}, 2005, pp. 539--546.

\bibitem{neff2019IOT}
C.~Neff, M.~Mendieta, S.~Mohan, M.~Baharani, S.~Rogers, and H.~Tabkhi,
  ``{REVAMP$^2$T}: Real-time edge video analytics for multi-camera
  privacy-aware pedestrian tracking,'' \emph{IEEE Internet of Things Journal},
  vol.~7, no.~4, pp. 2591--2602, 2020.

\bibitem{Denby2020ASPLOS_OEC}
B.~Denby and B.~Lucia, ``Orbital edge computing: Nanosatellite constellations
  as a new class of computer system,'' in \emph{Proceedings of the Twenty-Fifth
  International Conference on Architectural Support for Programming Languages
  and Operating Systems (ASPLOS)}, 2020, pp. 939--954.

\end{thebibliography}

\end{document}